\documentclass{article}
\usepackage{ijcai24}

\usepackage{times}
\usepackage{soul}
\usepackage{url}
\usepackage[hidelinks]{hyperref}
\usepackage[utf8]{inputenc}
\usepackage[small]{caption}
\usepackage{graphicx}
\usepackage{amsmath}
\usepackage{amsthm}
\usepackage{booktabs}
\usepackage{algorithm}
\usepackage{algorithmic}
\usepackage[switch]{lineno}

\usepackage{xcolor}
\definecolor{mypink}{rgb}{0.95, 0.5, 0.7}

\usepackage{hyperref}

\hypersetup{
    colorlinks=true,
    linkcolor=black,
    filecolor=black,      
    urlcolor=black,
    citecolor=black
}  

\usepackage[nameinlink]{cleveref}
\crefname{figure}{Fig.}{Figures}
\crefname{table}{Table}{Tables}
\crefname{appendix}{Appendix}{Appendixes}
\usepackage[utf8]{inputenc}
\usepackage[T1]{fontenc}    
\usepackage{makecell}       
\usepackage{subfigure}
\usepackage{tabularray}
\newcommand{\ourmodel}{ScreenAgent}

\usepackage{listings}
\usepackage{xcolor}
\usepackage{soul}
\usepackage{tcolorbox}

\title{\ourmodel~\includegraphics[width=0.06\textwidth]{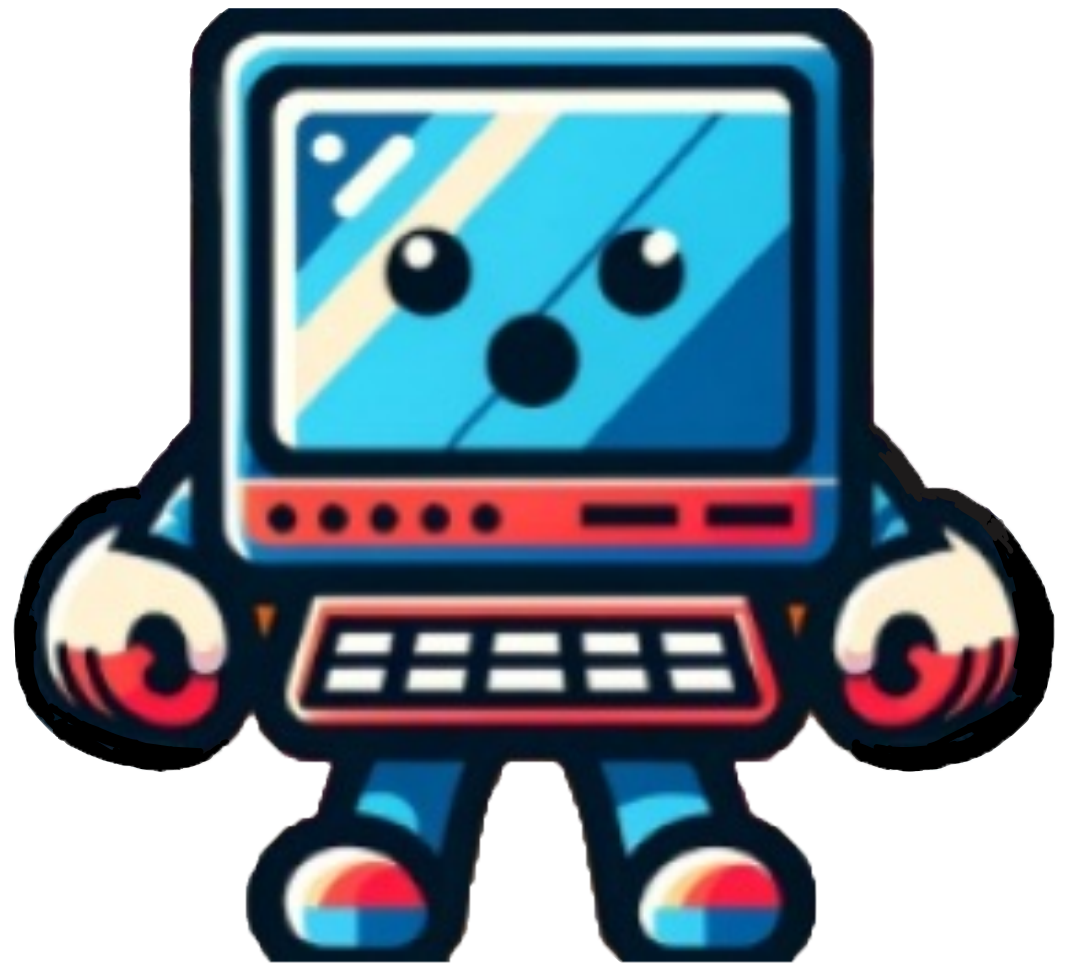}: A Vision Language Model-driven Computer Control Agent}

\author{
Runliang Niu$^{1}$\and
Jindong Li$^{1}$\and
Shiqi Wang$^{1}$\and
Yali Fu$^{1}$\and
Xiyu Hu$^{1}$\and\\
Xueyuan Leng$^{1}$\and
He Kong$^{1}$\and 
Yi Chang$^{1,2}$\and
Qi Wang$^{1,2}$\footnote{corresponding author}
\affiliations
$^1$ School of Artificial Intelligence, Jilin University\\
$^2$ Engineering Research Center of Knowledge-Driven Human-Machine Intelligence, \\ Ministry of Education, China\\
\footnote{\{niurl19,jdLi21,shiqiw23,fuyl23,xyhu23,lengxy22,\\konghe19\}@mails.jlu.edu.cn,
\{yichang,qiwang\}@jlu.edu.cn}
}

\begin{document}

\maketitle

\begin{abstract}
Existing Large Language Models (LLM) can invoke a variety of tools and APIs to complete complex tasks. The computer, as the most powerful and universal tool, could potentially be controlled directly by a trained LLM agent. Powered by the computer, we can hopefully build a more generalized agent to assist humans in various daily digital works. In this paper, we construct an environment for a Vision Language Model (VLM) agent to interact with a real computer screen. Within this environment, the agent can observe screenshots and manipulate the Graphics User Interface (GUI) by outputting mouse and keyboard actions. We also design an automated control pipeline that includes planning, acting, and reflecting phases, guiding the agent to continuously interact with the environment and complete multi-step tasks. Additionally, we construct the ScreenAgent Dataset, which collects screenshots and action sequences when completing a variety of daily computer tasks. Finally, we trained a model, ScreenAgent, which achieved computer control capabilities comparable to GPT-4V and demonstrated more precise UI positioning capabilities. Our attempts could inspire further research on building a generalist LLM agent. The code is available at \url{https://github.com/niuzaisheng/ScreenAgent}.
\end{abstract}

\section{Introduction}

\begin{figure}[t]
    \centering
    \includegraphics[width=0.5\textwidth]{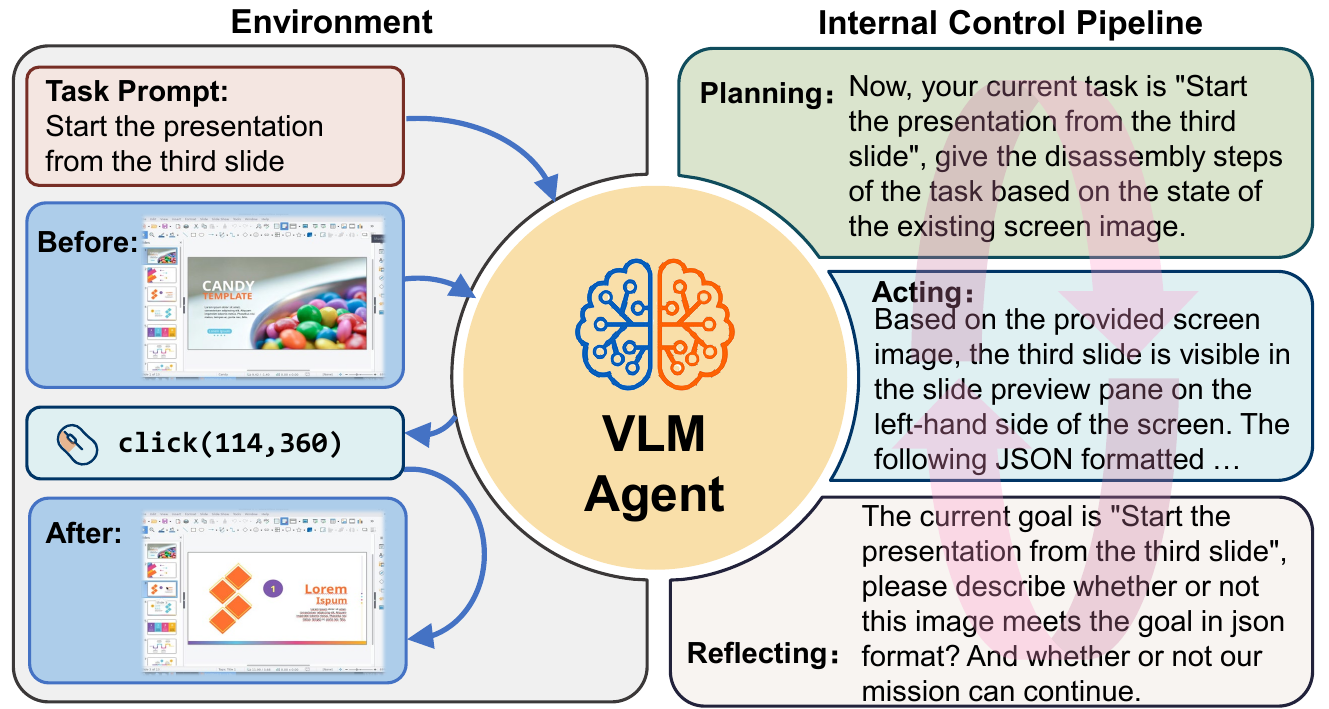}
    \caption{We have constructed a realistic computer-controlled environment and designed a control pipeline for the agent. The VLM agent retrieves instruction prompts and real computer states from the environment, then runs its internal control flow, going through the planning, acting, and reflecting phases. It outputs the next action operation, utilizes function calls to perform actions, induces changes in the computer environment, and achieves genuine real-time interaction between the agent and the environment.}
    \label{fig:Motivation}
    \vspace{-1em}
\end{figure}
Large language models (LLM), such as ChatGPT and GPT-4, have recently demonstrated exceptional performance in natural language processing tasks like generation, understanding, and dialogue. They have also significantly revolutionized research in other artificial intelligence fields. 
In particular, the development of these technologies paves the way for the study of intelligent LLM agents, which are capable of performing complex tasks~\cite{2023_arXiv_LLM_Survey}. A LLM agent is an AI entity with a large language model as its core computational engine. It possesses capabilities like Perception, Cognition, Memory, and Action, enabling the agent to perform highly proactive autonomous behaviors \cite{wang2023survey}. In LLM agent-related research, how to enable agents to learn to effectively use tools for expanding their action space has drawn extensive attention.



With the growing prevalence of electronic devices, such as personal computers, smartphones, tablets, and smart electronic instruments, our lives are becoming more intertwined with the digital world. Daily activities often require frequent interaction with electronic device screens. If an agent can free people from manual operations, and seamlessly navigate these devices by controlling screens according to user needs, it would mark a significant step towards achieving more general and autonomous intelligence~\cite{2023_arXiv_MLLM_Survey}. Indeed, a screen interaction agent must possess powerful visual information processing capabilities, and the ability to execute computer control instructions as shown in \cref{fig:Motivation}. Therefore, to achieve such a goal, it is necessary to first create a real interactive environment for the VLM agent, then guide the model and environment to form a continuous interactive pipeline, and train the agent to improve its performance. However, it's highly challenging to implement these functions within a unified framework and achieve satisfactory results from both the project engineering and theoretical research perspectives.

Despite the recent progress achieved by several works, some aspects still need to be further explored. For instance, CogAgent~\cite{2023_arXiv_CogAgent} specializes in GUI understanding and planning, showcasing remarkable proficiency in addressing diverse cross-modal challenges. However, CogAgent lacks the capability of a complete thinking chain, preventing it from forming a comprehensive tool invocation similar to ChatGLM~\cite{2022_ACL_ChatGLM,2022_arXiv_GLM-130B}. Later, AppAgent~\cite{2023_arXiv_AppAgent} focuses on smartphone operations, learning navigation, and acquiring new application usage through autonomous exploration or by observing human demonstrations. While AppAgent lacks a planning process and sacrifices the freedom of operation. It guides the agent to click by labeling each element, thereby limiting the method of tapping. As a result, current Vision Language Models (VLM agents) are typically unable to interact with real computer or mobile environments to generate and execute continuous manipulative commands.



To address the above-mentioned issues, we propose ScreenAgent, an entirely automated agent designed to handle continuous screen operations. This agent is primarily achieved through three components, namely planning, execution, and reflection. In particular, the reflecting module is inspired by Kolb's renowned experiential Learning Cycle theory~\cite{kolb_Learning_Cycle}, which enables the agent to perform reflective behaviors, making the entire pipeline more comprehensive and aligned with human action and thought processes. It autonomously assesses the execution status of the current action, providing feedback based on the ongoing state. This capability enhances its performance for subsequent actions, enabling our agent to possess the capability of a continuous thinking chain. Consequently, our agent can understand the next steps and engage in complete tool invocation to execute a series of continuous manipulative commands. The major contributions are summarized as follows:

\begin{itemize}
     \item We present a Reinforcement Learning (RL) environment that enables the VLM agent to directly interact with a real computer screen via VNC protocol. By observing the screenshot, our agent can interact with the GUI through basic mouse and keyboard operations.
    \item We develop an automated pipeline that encompasses the planning phase, acting phase, and reflecting phase. This integrated pipeline facilitates the agent's continuous interaction with the environment, distinguishing our agent from others.
    \item We propose the ScreenAgent dataset, which includes action sequences for completing generic tasks on Linux and Windows desktops. Moreover, we provide a fine-grained scoring metric to comprehensively evaluate the various capabilities that are necessary for a VLM agent in computer-controlling tasks.
    \item We test GPT-4V and two state-of-the-art open-source VLMs on our test set. The results demonstrate that GPT-4V is capable of controlling computers, but it lacks precise positioning capabilities. We thus trained a ScreenAgent to enable precise positioning and achieved comparable results to GPT-4V in all aspects. Our work can facilitate further research on building a generalist agent.
\end{itemize}

\section{Related Work}
\subsection{Multimodal Large Language Models}
LLMs have demonstrated powerful contextual understanding and text generation capabilities, enabling the implementation of complex multi-turn question-answering systems.
%
LLaMA~\cite{2023_arXiv_LLaMA} is a series of foundational language models spanning from 7 billion to 65 billion parameters, 
with Vicuna-13B~\cite{2023_Vicuna}, an open-source chatbot, being refined through fine-tuning on the LLaMA architecture.
GPT-4 is an advancement by OpenAI following the success of GPT-3 and it introduces several noteworthy improvements. 
GPT-4V(ision)~\cite{2023_arXiv_GPT-4V}, building upon GPT-4, has added multimodal capabilities.
LLaVA~\cite{2023_arXiv_LLaVA} and LLaVA-1.5~\cite{liu2023improved} connect the pre-trained CLIP~\cite{2021_ICML_CLIP} visual encoder with the Vicuna, achieving multimodal capabilities. 
Fuyu-8B\footnote{\url{https://www.adept.ai/blog/fuyu-8b}} does not use an image encoder but opts for a pure decoder Transformer architecture.
CogVLM~\cite{2023_arXiv_CogVLM} is a powerful open-source Visual Language Model that supports image understanding at a resolution of 490 × 490 and multi-turn dialogues.
In addition, Monkey~\cite{2023_arXiv_Monkey} introduces an efficient training method that enhances input resolution capability.


\begin{figure*}[t!]
    \centering
    \includegraphics[width=1\textwidth]{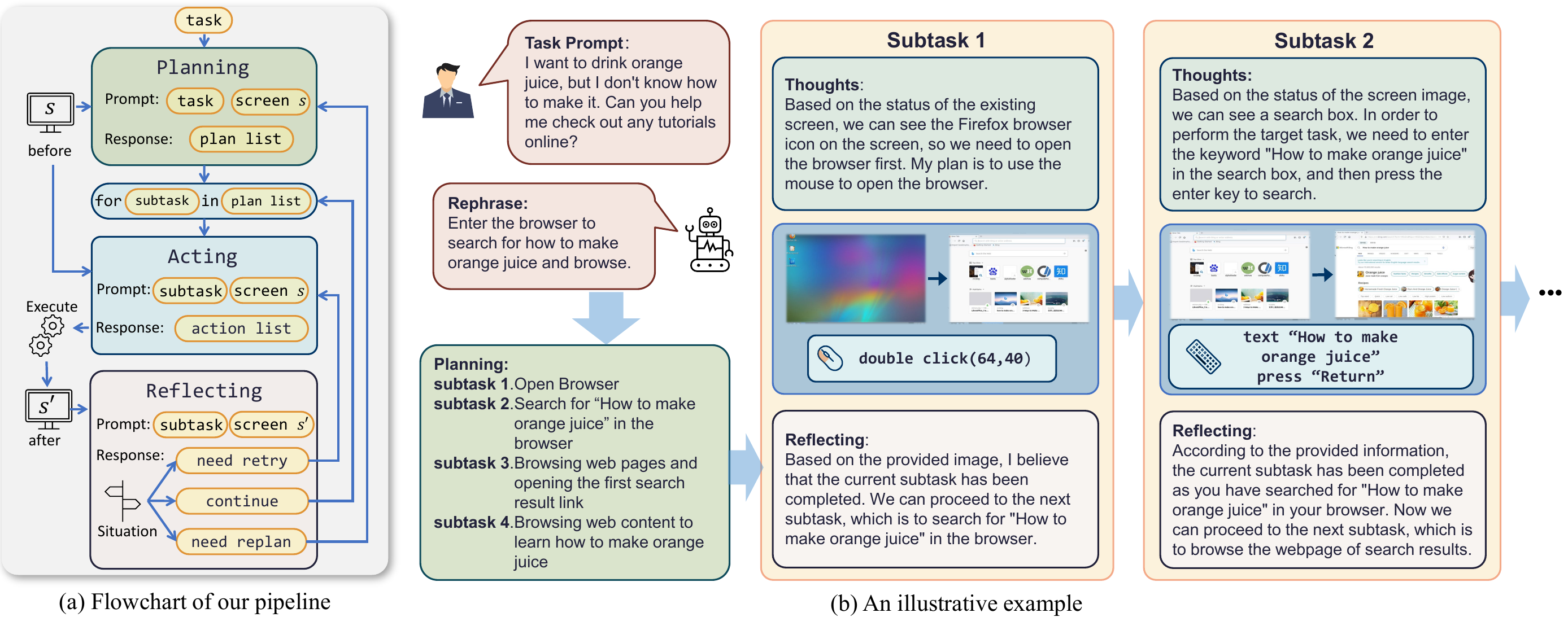}
    \caption{The overview of our computer control pipeline, which includes planning, acting, and reflecting phases. Sub-figure (a) presents the flowchart of our pipeline, while sub-figure (b) provides an illustrative example.
    Based on the user's task prompt, the agent initially decomposes the task into subtasks.
    In each subtask, the agent first describes the screen and generates mouse and keyboard operations in a function-call style.
    In the reflection phase, the agent decides whether to proceed to the next subtask, retry the current subtask, or reformulate the entire plan.}
    \label{fig:Pipeline}
\end{figure*}

\subsection{Computer Control Environment \& Dataset}
In simulated environments, agents can be trained to emulate clicking and typing. 
WebNav~\cite{nogueira2016end} creates a navigation environment with links, testing the agent's sequential decision-making ability.
MiniWoB++ ~\cite{liu2018reinforcement} provides a lot of simplified ATARI-like web-browser-based tasks as a Reinforcement Learning (RL) environment. 
WebShop~\cite{yao2023webshop} offers tasks for controlling the browser to complete the purchase process. 
SWDE~\cite{haoOneTreeForest2011} preserves webpage HTML files to train information extraction models.
WebSRC~\cite{chenWebSRCDatasetWebBased2021} is a QA-style dataset that contains a large number of questions and answers about webpage screenshots. 
Mind2Web~\cite{dengMind2WebGeneralistAgent2023} introduces a dataset for developing generalist web agents. 
Seq2act~\cite{seq2act2020Li} integrates three datasets for Android, AndroidHowTo, Rico-SCA, and PixelHelp. 
Screen2Words~\cite{Wang2021Screen2Words} is a large-scale screen summarization dataset for Android UI screens.
META-GUI~\cite{sun2022meta} is a dataset for training a multi-modal conversational agent on mobile GUI.
~\cite{burns2022motifvln} provides a dataset of unknown command feasibility on Android. 

\subsection{Large Language Model-Driven Agents}
With the advancement of large language models, the capabilities of intelligent agents have also been enhanced.
WebGPT~\cite{2021_arXiv_WebGPT} conducts fine-tuning on GPT-3 to address extended questions within a text-based web-browsing environment, enabling the model to explore and navigate the web for answers.
ToolFormer~\cite{2023_arXiv_ToolFormer} integrates an assortment of utilities, featuring a calculator, Q\&A system, and search engine, among others.
Voyager~\cite{2023_arXiv_Voyager} stands as the inaugural embodiment of a Large Language Model-powered, lifelong learning agent within the Minecraft environment. 
RecAgent~\cite{2023_arXiv_RecAgent} proposes that agents can generate high-level thoughts through the operation of memory reflection.
ProAgent~\cite{2023_arXiv_ProAgent} introduces a novel paradigm in process automation that seamlessly integrates agents powered by Large Language Models. 
CogAgent~\cite{2023_arXiv_CogAgent}, an 18-billion-parameter visual language model, is meticulously designed for GUI comprehension and navigation. 
AppAgent~\cite{2023_arXiv_AppAgent} proposes a multi-modal agent framework for learning operations on smartphone applications.

\section{Framework}
In this section, we introduce our Reinforcement Learning (RL) environment and the autonomous control flow within the Agent. Through this environment, a VLM agent can interact with a real computer screen, observe screen images, select actions, and autonomously complete specific tasks.

\subsection{Computer Control Environment}
We construct a computer control environment to assess the capabilities of VLM agents. This environment connects to a desktop operating system through remote desktop (VNC) protocol and allows the sending of mouse and keyboard events to the controlled desktop. The formal definitions of this environment are defined as follows: 
\begin{itemize}
    \item \textbf{$A$-Action Space.} We define an action as a form of a function call. If the agent outputs an action in the required JSON-style format, the action will be parsed and executed by the environment. All action types and corresponding action attributes are defined in \cref{tab:action_space}.
    \item \textbf{$S$-State Space.} The screenshot image is utilized as the state space of the environment. The environment will collect screenshots $s$ and $s'$, denoting the state before and after each action, respectively.
    \item \textbf{$R$-Reward Function.} Due to the highly open-ended nature of the task, the reward function in this environment is flexibly opened to different interfaces, which can integrate different existing or future reward models.
\end{itemize}

Through remote control, the agent can perform arbitrary tasks on the screen, which creates a highly challenging open environment having a large state and action space.

\begin{table}[t]
    \centering
    \scalebox{0.8}{
        \begin{tblr}{
          column{1} = {c},
          cell{1}{1} = {c=2}{},
          cell{2}{1} = {r=6}{},
          cell{8}{1} = {r=2}{},
          cell{10}{1} = {c=2}{},
          cell{11}{1} = {c=2}{},
          cell{12}{1} = {c=2}{},
        }
        \hline
        Action Type &  & Attributes\\\hline
        Mouse & Move & Mouse Position(width:int, height:int)\\\hline
         & Click & \makecell[l]{Mouse Button(left/middle/right), \\ Mouse Position(width:int, height:int)}\\\hline
         & Double Click & \makecell[l]{Mouse Button(left/middle/right), \\Mouse Position(width:int, height:int)}\\\hline
         & Scroll Up & Scroll Repeat(int)\\\hline
         & Scroll Down & Scroll Repeat(int)\\\hline
         & Drag & Drag End Position(width:int, height:int)\\\hline
        Keyboard & Press & Keyboard Key or Combined-keys (string)\\\hline
         & Text & Keyboard Text(string)\\\hline
        Wait Action &  & Wait Time(float) \\\hline
        Plan Action &  & Element(string) \\\hline
        Evaluate Action &  & \makecell[l]{Situation(success/retry/reformulate) \\ Advice(string) } \\\hline
        \end{tblr}
    }
    \caption{All supported action types and action attributes.}\label{tab:action_space}
    \vspace{-1em}
\end{table}

\subsection{Control Pipeline}
To guide the agent to continually interact with the environment and complete multi-step complex tasks. We designed a control pipeline including the Planning, Acting, and Reflecting phases. The whole pipeline is depicted in ~\cref{fig:Pipeline}.
The pipeline will ask the agent to disassemble the complex task, execute subtasks, and evaluate execution results. The agent will have the opportunity to retry some subtasks or adjust previously established plans to accommodate the current occurrences. The details are depicted as follows:

\textbf{Planning Phase.} In the planning phase, based on the current screenshot, the agent needs to decompose the complex task relying on its own common-sense knowledge and computer knowledge.

\textbf{Acting Phase.} 
In the acting phase, based on the current screenshot, the agent generates low-level mouse or keyboard actions in JSON-style function calls.
The environment will attempt to parse the function calls from the agent's response, and convert them to device actions defined in the VNC protocol. Then our environment will send actions to the controlled computer. The environment will capture the after-action screen as input for the next execution phase.

\textbf{Reflecting Phase.} 
The reflecting stage requires the agent to assess the current situation based on the after-action screen. The agent determines whether needs to retry the current sub-task, go on to the next sub-task, or make some adjustments to the plan list.
This phase is crucial within the control pipeline, providing some flexibility to handle a variety of unpredictable circumstances.


All prompts and templates in the pipeline are provided in \cref{appendix:Agent Prompt Details}. 

\begin{figure}[t]
    \centering
    \includegraphics[width=\linewidth]{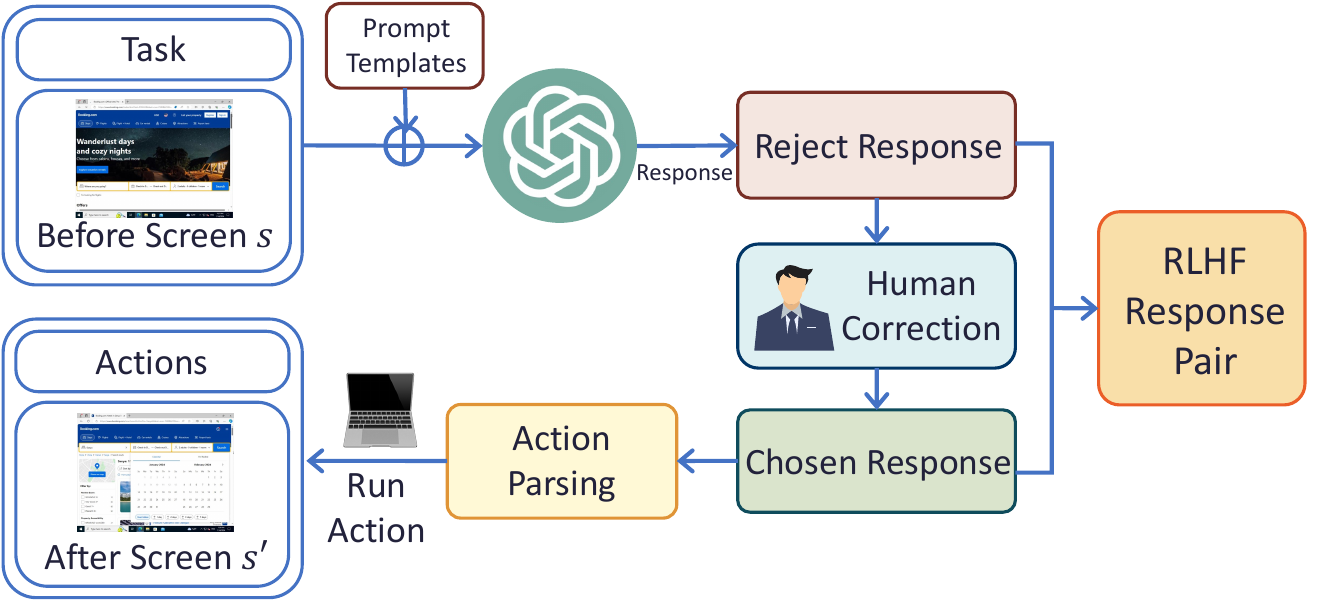}
    \caption{Data annotation process.
    We invoke GPT-4V to generate an original response, and annotators correct this response as the golden labeled response. The environment parses executable actions from the text and sends them for real computer execution.
    The original response and the golden labeled response form a pair, which can be utilized for training in future Reinforcement Learning from Human Feedback (RLHF) processes.
    }
    \label{fig:InteractiveAnnotationProcess}
\end{figure}

\begin{figure}[h]
    \centering
    \includegraphics[width=0.48\textwidth]{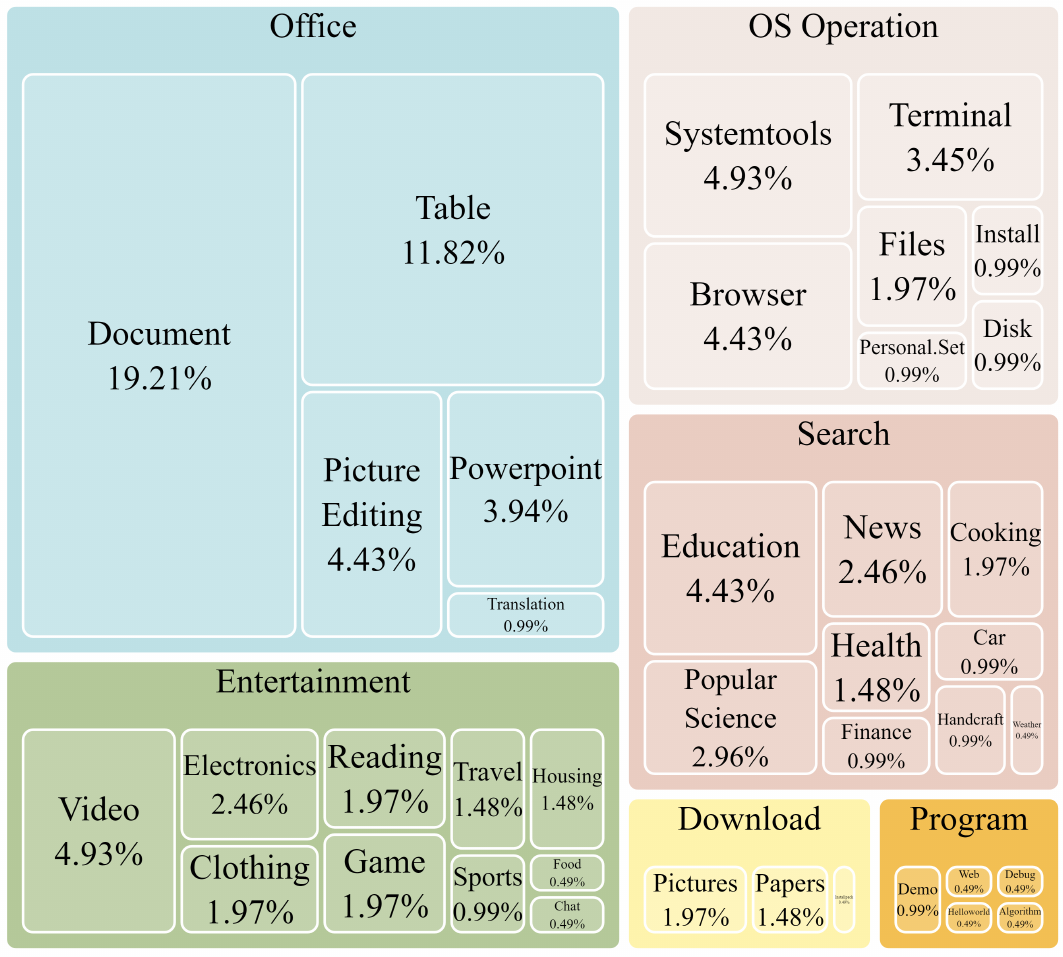}
    \caption{Task type statistics in ScreenAgent training set.}
    \label{fig:Screen Agent Dataset_Treemap}
    \vspace{-1em}
\end{figure}

\begin{figure}[t]
    \centering
        \subfigure[~]{
            \includegraphics[width=0.14\textwidth]{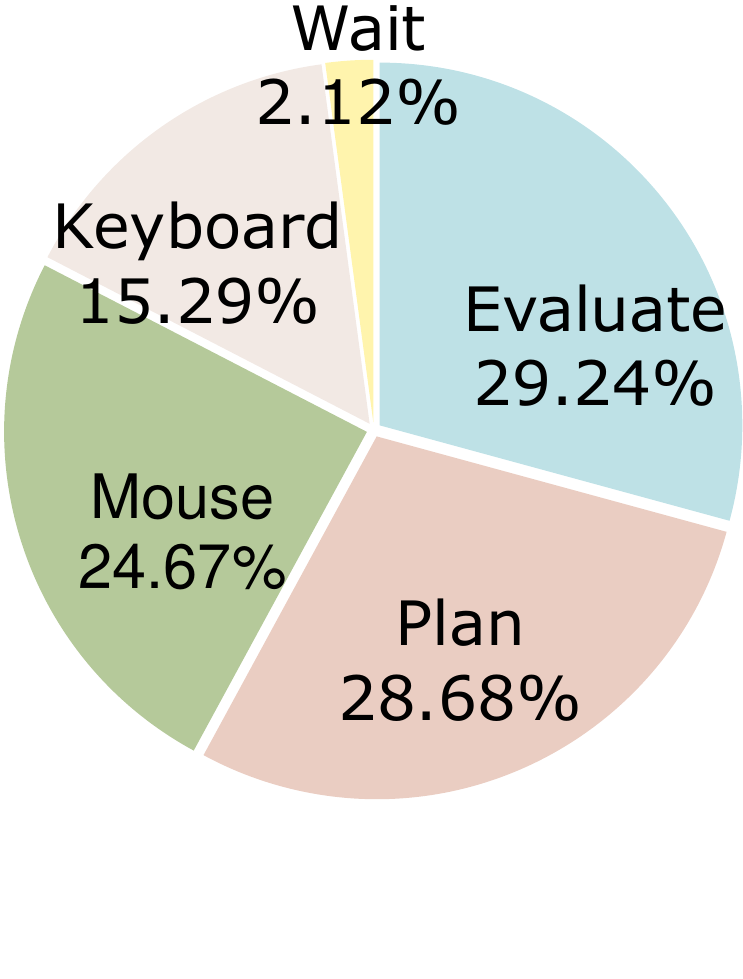}
            \label{fig:Screen Agent Dataset_(b)}
            }
        \subfigure[~]{
            \includegraphics[width=0.27\textwidth]{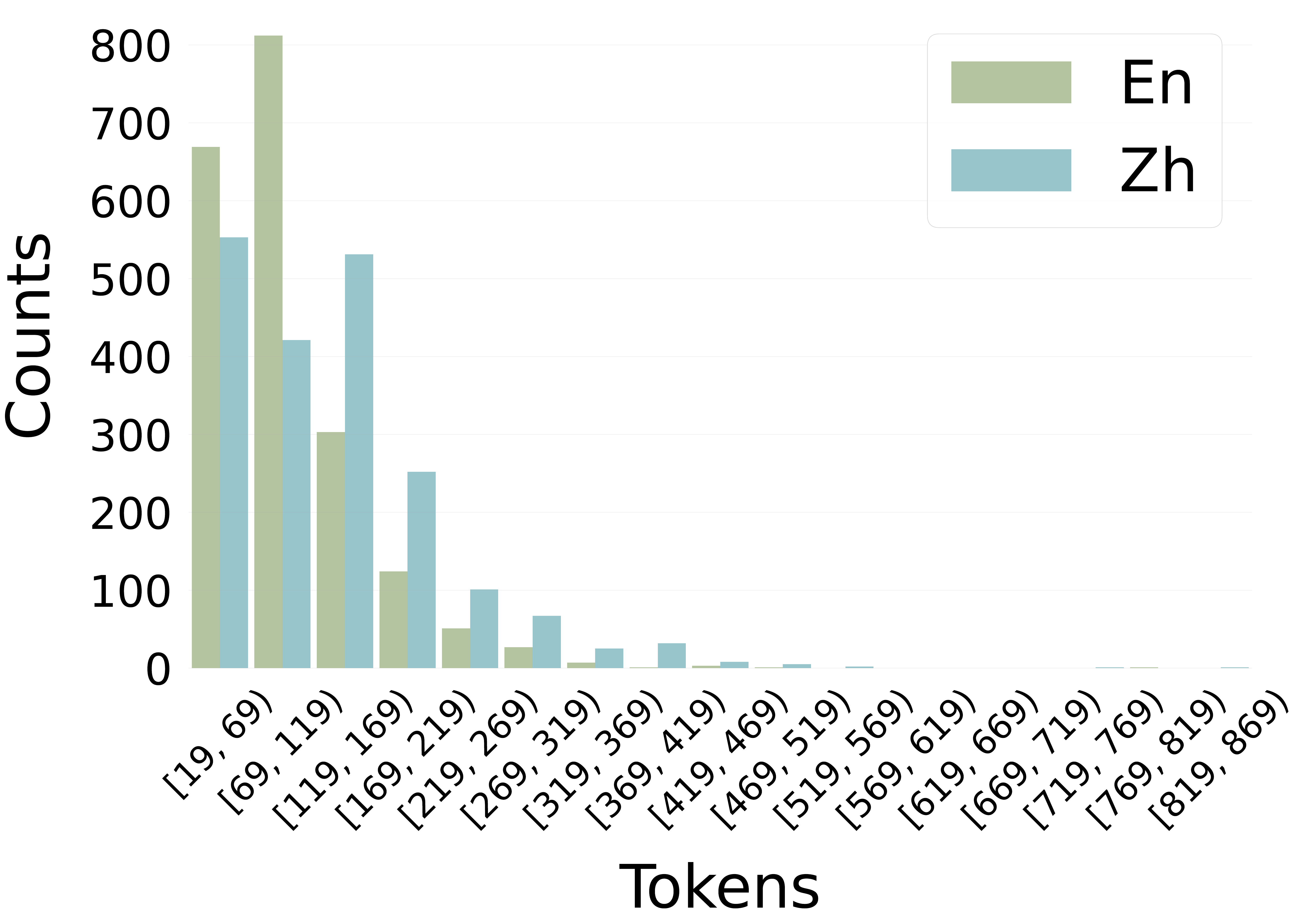}
            \label{fig:Screen Agent Dataset_(d)}
            }
    \caption{The statistical information of ScreenAgent training set: (a) Distribution of action types; (b) Chosen response token number distribution.}
    \label{fig:ScreenAgentDataset}
    \vspace{-1em}
\end{figure}


 \begin{table*}[th]
    \centering
    \renewcommand{\arraystretch}{1.3}
    \scalebox{0.9}{%
        \begin{tabular}{c|ccccccc}
            \toprule
            \makecell[c]{\textbf{Model}} & 
            \makecell[c]{\textbf{Plan} \\ total 284 } & 
            \makecell[c]{\textbf{Action} \\ \textbf{Type} \\ total 650} & 
            \makecell[c]{\textbf{Mouse} \\ \textbf{Action} \\ \textbf{Type} \\ total 232 } & 
            \makecell[c]{\textbf{Mouse} \\ \textbf{Button} \\ total 209 } &
            \makecell[c]{\textbf{Mouse} \\ \textbf{Position} \\ total 218 } &
            \makecell[c]{\textbf{Keyboard} \\ \textbf{Keys} \\ \textbf{or Text} \\ total 134} & 
            \makecell[c]{\textbf{Reflecting} \\ \textbf{Situation} \\ \textbf{Assessment} \\ total 546} \\
            \midrule
            \textbf{ LLaVA-1.5 }  & \underline{0.78} & 0.75 & 0.71 & 0.74 & 0.72 & 0.45 & 0.98 \\
            \textbf{CogAgent-VQA} & 0.00 & 0.03 & 0.06 & 0.06 & 0.05 & 0.01 & 0.39 \\
            \makecell[c]{\textbf{CogAgent-Chat} (original output)}
                                  &  0.00 &  0.00  &  0.00  &  0.00  &  0.00  &  0.00  & 0.30  \\
            \makecell[c]{\textbf{CogAgent-Chat} (help by GPT-3.5)} 
                                  &0.29 &  0.38 & 0.44 & 0.45 & 0.42 & 0.17 & 0.76 \\
            \textbf{GPT-4V(ision)}& \textbf{0.87} & \textbf{0.86} & \underline{0.85} & \underline{0.85} & \underline{0.83} & \underline{0.77} & \textbf{1.00} \\
            \midrule
            \textbf{ScreenAgent}  & 0.72 & \underline{0.83} & \textbf{0.91} & \textbf{0.92} & \textbf{0.91} & \textbf{0.82} & \textbf{1.00}\\
            \bottomrule
        \end{tabular}%
    }
    \caption{Proportion of successful function call on ScreenAgent test-set.}
    \label{tab:follow_instruction_result}
    \vspace{-1em}
\end{table*}

 \begin{table*}[t]
    \centering
    \renewcommand{\arraystretch}{1.3}
    \scalebox{0.83}{%
        \begin{tabular}{c|c|ccccccc}
            \toprule
            \makecell[c]{\textbf{Model}} & 
            \makecell[c]{\textbf{CC-Score}} & 
            \makecell[c]{\textbf{Plan} \\ (BLEU) } & 
            \makecell[c]{\textbf{Action} \\ \textbf{Type} \\ (F1)} & 
            \makecell[c]{\textbf{Mouse} \\ \textbf{Action} \\ \textbf{Type} \\ (F1) } & 
            \makecell[c]{\textbf{Mouse} \\ \textbf{Button} \\ (F1) } &
            \makecell[c]{\textbf{Mouse} \\ \textbf{Position} \\ (Accuracy)} &
            \makecell[c]{\textbf{Keyboard} \\ \textbf{Keys} \\ \textbf{or Text} \\ (BLEU)} & 
            \makecell[c]{\textbf{Reflecting} \\ \textbf{Situation} \\ \textbf{Assessment}\\ (F1)} \\
            \midrule
            \textbf{ LLaVA-1.5 }  & 0.51 & 0.29 & 0.91 & 0.90 & 0.96 & 0.03 & 0.70 & \underline{0.52} \\
            \makecell[c]{\textbf{CogAgent-Chat} (help by GPT-3.5)} 
                                  &  0.33 & \underline{0.32} & 0.83 & 0.86 & 0.02 & 0.07 & 0.74 & 0.51  \\
            \textbf{GPT-4V(ision)}& \textbf{0.63} & \textbf{0.47} & \textbf{0.98} & \textbf{0.96} & \textbf{0.99} & \underline{0.12} & \textbf{0.92} & \textbf{0.60} \\
            \midrule
            \textbf{ScreenAgent}  & \underline{0.61} & 0.31 & \textbf{0.98} & \underline{0.94} & \underline{0.97} & \textbf{0.51} & \underline{0.87} & \underline{0.52} \\
            \bottomrule
        \end{tabular}%
    }
    \caption{Comparison of VLM fine-grained score in all successful matched action on ScreenAgent Test-Set.}
    \label{tab:fine_grained_result}
\end{table*}

\section{ScreenAgent Dataset \& CC-Score}
Existing computer-controlled datasets typically have a narrow range of applicability scenarios.
For instance, building upon the foundational premise that it is easy to obtain UI element metadata through HTML or developer modes, WebNav~\cite{nogueira2016end}, Mind2Web~\cite{dengMind2WebGeneralistAgent2023}, and SWDE~\cite{haoOneTreeForest2011} mainly focused on web browsing, while Seq2act~\cite{seq2act2020Li} and Screen2Words~\cite{Wang2021Screen2Words} are tailored for Android. 
However, the mouse and keyboard are also common and universal interfaces to control a computer. To fill the gap in this type of control method, we build an interactive annotation process (shown in~\cref{fig:InteractiveAnnotationProcess}) to construct the ScreenAgent Dataset which is collected from Linux and Windows operating systems for completing specific tasks. This dataset endeavors to cover a wide range of daily computer usage scenarios, including daily office, booking, information retrieval, card games, entertainment, programming, system operations, and so on.
As illustrated in \cref{fig:Screen Agent Dataset_Treemap}, the ScreenAgent Dataset encompasses 39 sub-task categories across 6 themes. The dataset has 273 complete task sessions, with 203 sessions (3005 screenshots) for training and 70 sessions (898 screenshots) for testing. \cref{fig:ScreenAgentDataset} shows important statistical information about the dataset. More statistical information and samples are provided in \cref{appendix:dataset_detail}.

To assess an agent's capability in the computer control task, we have designed a fine-grained evaluation metric Vision Language Computer Control Score (CC-Score) for assessing the similarity of control action sequences. This metric takes into account both the sequential order and actions' attribution.
We developed specific similarity metrics for every action type.
For mouse actions, the metrics include four aspects of consistency: action type, mouse operation type, mouse button, and whether the click coordinates are within the annotated feasible bounding box.
For text and keyboard actions, the metrics involve two aspects: action type consistency and the BLEU score of the input text, single key, or keyboard shortcut combination.
For the entire action sequence, we employ an alignment algorithm that identifies the maximum matching pairs of predicted action and labeled action, while maintaining the sequence order. This approach maximizes the overall score, which is used as the measure of sequence similarity. 
Ultimately, the CC-Score encompasses the normalized scores of predicted and labeled sequences, the F1 values for each action attribute classification, and the unigram similarity values for text types. The implementation details of CC-Score are provided in \cref{appendix:CC-Score}.

\section{Experiment}

In the experimental phase, we assessed OpenAI GPT-4V performance on the ScreenAgent test set, along with evaluations of three open-source VLMs. Furthermore, one of these models underwent fine-tuning to potentially augment its proficiency in screen control tasks. Subsequently, we conducted a thorough analysis of the outcomes and identified several typical cases to elucidate the inherent challenges of our task.

\subsection{Evaluation Results on ScreenAgent Test-Set}

Apart from GPT-4V, we selected several recently released SoTA VLMs for testing, including LLaVA-1.5~\cite{liu2023improved} and CogAgent~\cite{2023_arXiv_CogAgent}. LLaVA-1.5 is a 13B-parameter multimodal model, unfortunately, it only supports up to 336 × 336px image size inputs. CogAgent is an 18B-parameter visual language model designed for GUI comprehension and navigation. Leveraging dual image encoders for both low-resolution and high-resolution inputs, CogAgent demonstrates proficiency at a resolution of 1120 × 1120px, allowing it to discern minute elements and text.

\begin{figure}[t]
    \centering
    \includegraphics[width=0.47\textwidth]{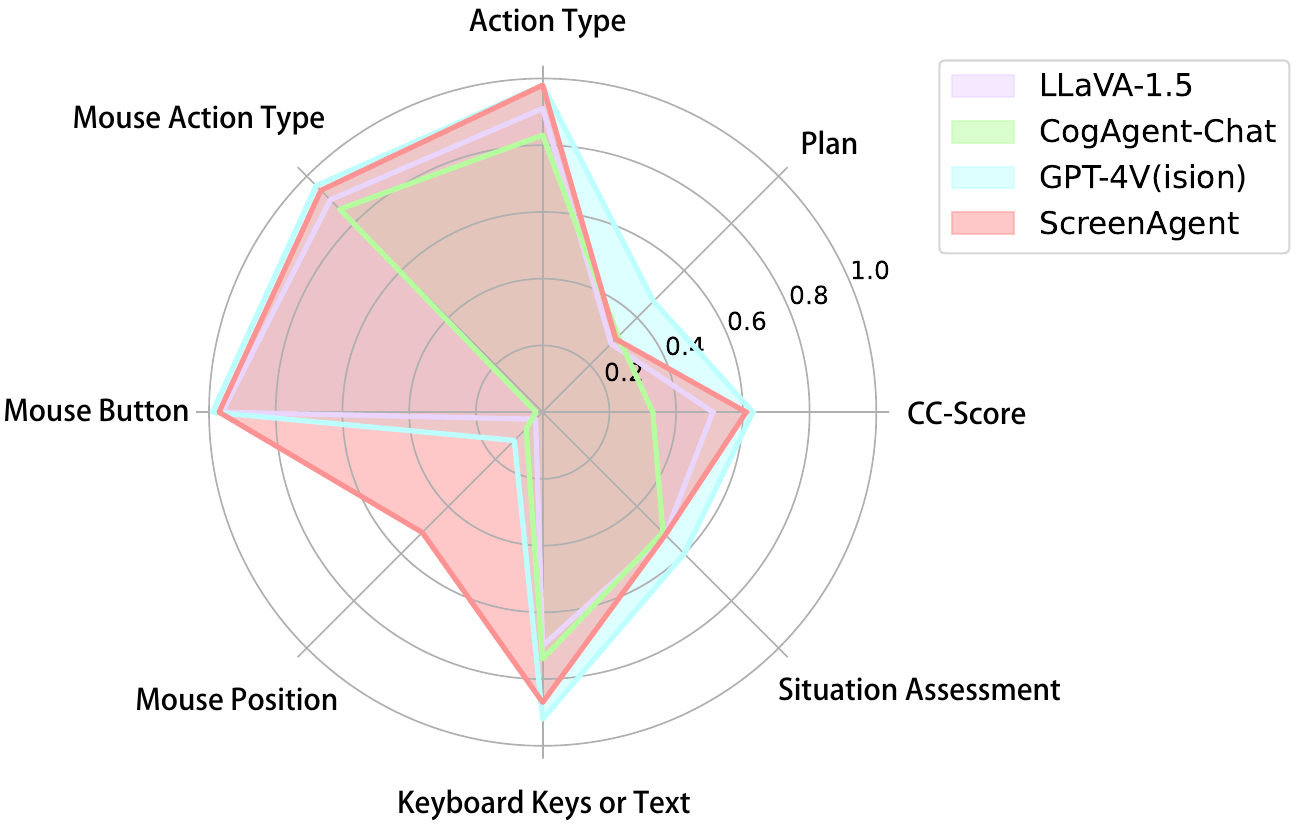}
    \caption{ScreenAgent can complete computer control tasks most excellently compared with other VLMs/Agents.}
    \label{fig:radar}
\end{figure}

We test the models' capabilities from two aspects: The ability to follow instructions to output the correct function call format, shown in \cref{tab:follow_instruction_result}, and the ability to complete specific tasks assigned by the user, shown in \cref{tab:fine_grained_result}.

Following instructions and executing correct function calls is the most fundamental skill for an LLM agent when learning to use external tools.
\cref{tab:follow_instruction_result} presents the success rate of these function calls for each attribute key. This assessment focuses on whether the model can accurately execute various functions encompassing the attribute items expected by manual action annotations. Note that, this evaluation does not consider the consistency of the attribute values with the golden labeling; it solely examines if the model's output includes the necessary attribute keys. From the table, GPT-4V and LLaVA-1.5 achieved higher scores, while CogAgent and its upstream model CogAgent-VQA underperformed. CogAgent-VQA and CogAgent-Chat almost entirely disregarded the JSON format action definitions in our prompts, resulting in a very low score on successful function calls. Therefore, rendering them completely incapable of interacting with our environment. To ensure fairness in comparison, we utilize OpenAI GPT-3.5 to extract action into JSON-style function calls from the original CogAgent-Chat responses, indicated as "CogAgent-Chat (helped by GPT-3.5)". Even so, its scores are significantly lower than those of LLaVA-1.5 and GPT-4V, although CogAgent has been trained on Mind2Web web browsing simulation datasets.

\begin{figure*}[h]
    \centering
    \includegraphics[width=1\textwidth]{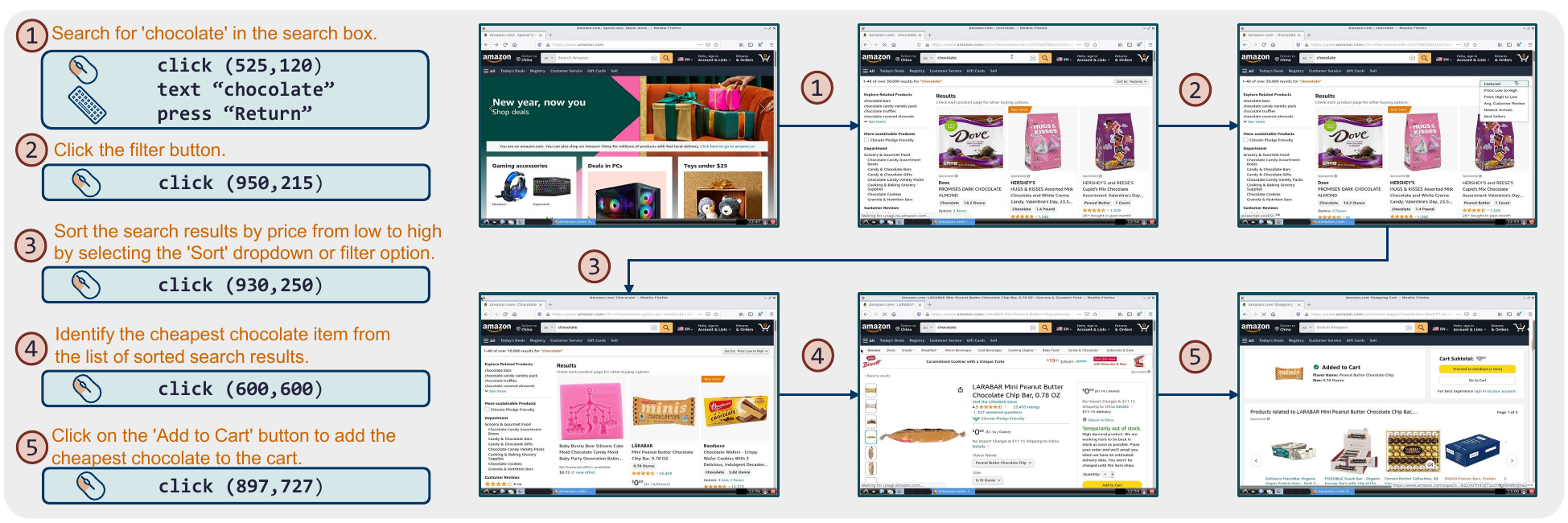}
    \caption{An example of the action task, where we assign ScreenAgent the following task: "Find and add the cheapest chocolate to the cart on Amazon". The diagram delineates the actions that the agent needs to perform, along with the alterations in the computer environment before and after the execution of these actions.
    }
    \label{fig:case_chocolate}
\end{figure*}

\cref{tab:fine_grained_result} displays the fine-grained scores of predicted attribute values for each action within the successfully parsed function calls. As can be seen, GPT-4V remains the best performer, with action type prediction F1 score of 0.98. This implies that it can accurately select appropriate mouse or keyboard actions. Additionally, it can precisely choose the mouse action type, typing text, or pressing keys consistent with the golden label actions. 

The ability for precise positioning is crucial in computer-controlling tasks. As indicated by the "Mouse Position" column in \cref{tab:fine_grained_result}, current VLMs have not yet achieved the capability for precise positioning required for computer manipulation. GPT-4V refuses to give precise coordinate results in its answers, and two open-source models also fail to output the correct coordinates with our pipeline prompt template.

Another significant challenge for all models is the reflection phase. In this phase, the agent is required to determine whether the subtask has been completed in the current state, and decide whether to proceed further or make some adjustments. This is crucial for constructing a continuous interactive process. Regrettably, all models show insufficient accuracy in this determination, with GPT-4V achieving only a 0.60 F1 score. This implies that human intervention is still necessary during task execution.

\begin{table}
    \centering
    \renewcommand{\arraystretch}{1.2}
    \scalebox{0.88}{
        \begin{tabular}{c|c|cccc}
          \toprule
          \textbf{Dataset}    &\textbf{Samples}& \textbf{1} & \textbf{2} & \textbf{3} & \textbf{4}\\ \hline
          \textbf{COCO}                & 42404 & 20\% & 10\% & -   & -   \\
          \textbf{Widget Captions}     & 41221 & 20\% & 10\% & -   & -   \\
          \textbf{Mind2Web}            & 12846 & 30\% & 40\% & 50\% & 30\% \\
          \textbf{ScreenAgent Dataset} & 3005  & 30\% & 40\% & 50\% & 70\% \\
          \bottomrule
        \end{tabular}
    }
    \caption{Training data proportions and division of four training phases. Percentages indicate the proportion of samples from this data set at each phase.}
    \label{tab:data_proportion}
    \vspace{-1em}
\end{table}

\begin{figure*}[t!]
    \centering
    \includegraphics[width=1\textwidth]{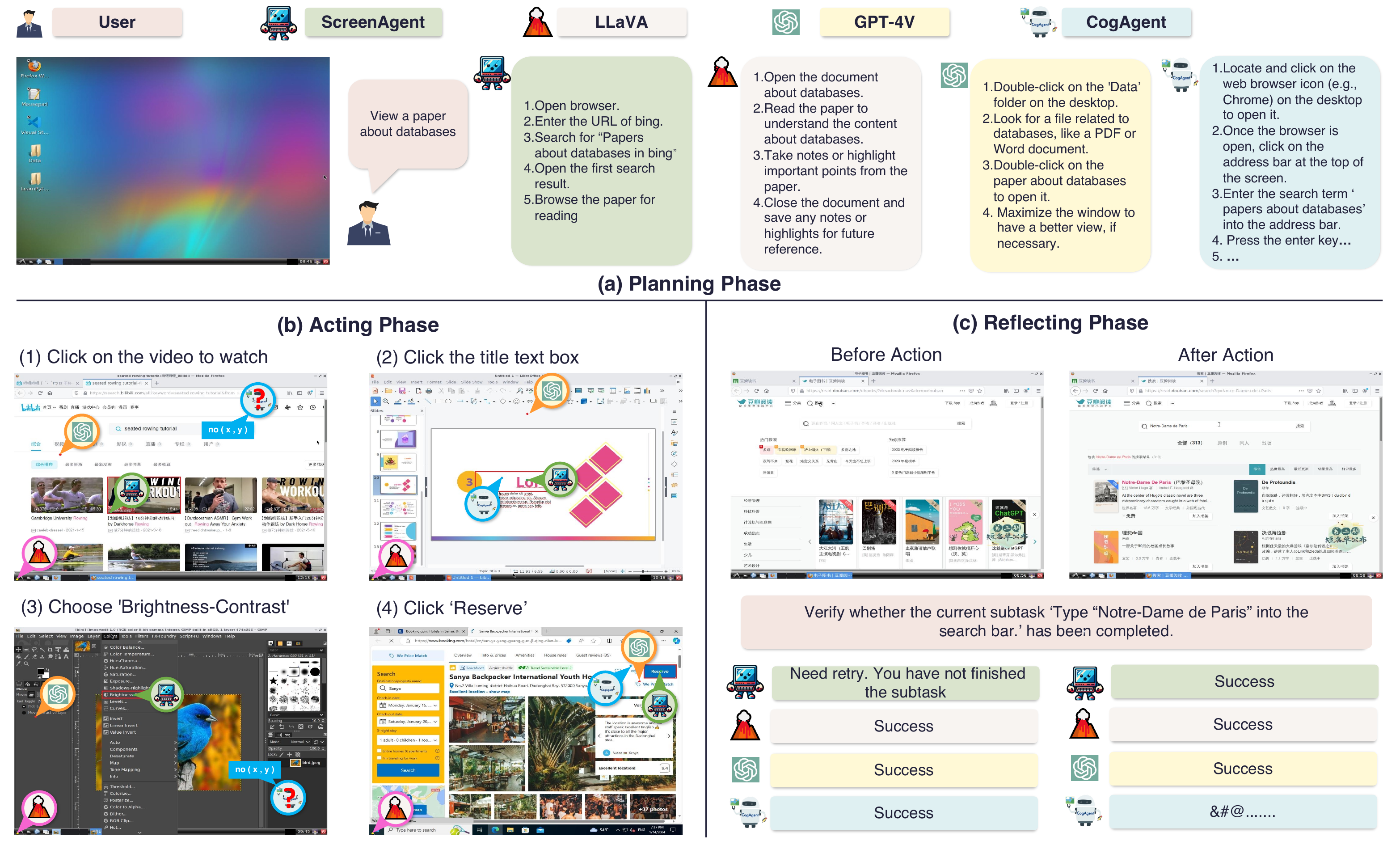}
    \caption{An example to show the execution results of multiple VLM agents among the Planning, Acting, and Reflecting phases in our pipeline.}
    \label{fig:case_comparison}
\end{figure*}

\subsection{Fine-tuning Training}
To demonstrate the potential for ongoing research on the task, we continue to fine-tune the CogAgent-Chat model on our ScreenAgent training set to enhance its function call ability. Similar to the approach adopted in recent VLM works~\cite{chen2023minigptv2}, we mix data from multiple datasets and construct four distinct training phases, which is illustrated in \cref{tab:data_proportion}.
We reformulated two objective detection datasets, COCO~\cite{lin2015microsoft} and Widget Captions \cite{li2020widget}, into mouse-click tasks to enhance the model's localization ability. For Mind2Web, we implemented a series of complex data augmentations to align with our task objectives. The details of the data construction are outlined in \cref{appendix:data_construction}. 

After vision fine-tuning, ScreenAgent achieved the same level of following instructions and making function calls as GPT-4V on our dataset, as shown in \cref{tab:follow_instruction_result}. In \cref{tab:fine_grained_result}, ScreenAgent also reached a comparable level to GPT-4V. Notably, our ScreenAgent far surpasses existing models in the precision of mouse clicking. This indicates that vision fine-tuning effectively enhances the model's precise positioning capabilities. Additionally, we observed that ScreenAgent has a significant gap compared to GPT-4V in terms of task planning, highlighting GPT-4V's common-sense knowledge and task-planning abilities.

\subsection{Case Study}
To evaluate our ScreenAgent model on computer control tasks, we provide two cases. 
In \cref{fig:case_chocolate}, we present a case illustrating the workflow of ScreenAgent executing a chain of actions. 
In \cref{fig:case_comparison}, we compare different agents in executing the details of the three phases in the pipeline. 
\cref{fig:case_comparison} (a) shows the planning process of all the agents, where we find that our ScreenAgent produces the most concise and effective plan.
\cref{fig:case_comparison} (b) presents four different click action tasks, each representing a step in a specific task. Results show that LLaVA clicks on the bottom-left corner on all screens, cogAgent may fail to generate click positions, and in the fourth task, only our agent can correctly click on the position.
\cref{fig:case_comparison} (c) shows that our agent can recognize whether an action needs to be re-tried after reflection and successfully execute the action following a failure.

\section{Conclusion}
In this paper, we build a new environment for the screen control task. VLM agents can manipulate a real computer through direct mouse and keyboard control commands.
To encourage the agent to continuously interact with the environment and accomplish complex multi-step tasks, we designed a control pipeline that includes planning, acting, and reflecting phases. 
Furthermore, we unveil a new dataset that covers a wide range of everyday digital works.
We propose a fine-grained metric to assess the agent's computer-controlling capabilities with both action-level and task-level evaluation.
We tested OpenAI GPT-4V and two state-of-the-art VLM models on the test set. The results indicate that GPT-4V has the potential to act as a computer-controlling agent, but it lacks precise positioning capabilities. 
Finally, we trained the ScreenAgent model, inherited from CogAgent, to achieve comparable scores with GPT-4V but more accuracy in UI element localization.
It is hoped that our work will inspire further research in building more powerful and generalized agents. In terms of technical limitations, due to the input restrictions of VLM, our model can only process single-frame images, not videos or multi-frame images. The VLM’s language capability is limited by the abilities of its foundation language model. We also found that even GPT-4V has limited support for non-English text on the screen.

\section*{Ethical Statement}
The rational use of automated agents with autonomous decision-making capabilities brings significant societal benefits, including improving the accessibility of computers, reducing duplication of human effort on digital work, and aiding in computer education. However, the potential negative impacts of these agents, such as employment impact, fraud and abuse, privacy issues, and the risk of misoperation, cannot be overlooked. 
The method could potentially be used to bypass the CAPTCHA test for automatic account registration, spreading misinformation, or conducting illegal activities.
We are focused on and committed to the responsible use of AI technology.



\bibliographystyle{named}
\bibliography{ijcai24}

\newpage

\appendix

\onecolumn

\section*{\centering \LARGE ScreenAgent \includegraphics[width=0.06\textwidth]{figs/ScreenAgent.png}: A Vision Language Model-driven Computer Control Agent}
\section*{\centering \LARGE \textmd{Appendix}}

\section{Agent Prompt Details}\label{appendix:Agent Prompt Details}

\definecolor{backcolour}{rgb}{0.95,0.95,0.92}

\lstdefinelanguage{json}{
    basicstyle=\normalfont\ttfamily,
    showstringspaces=false,
    breaklines=true,
    backgroundcolor=\color{backcolour},
    commentstyle=\color{gray}\ttfamily,
    keywordstyle=\color{blue}\ttfamily,
    stringstyle=\color{blue}\ttfamily,
    keywordstyle={[2]\color{blue}\ttfamily},
    morekeywords={[2] 
        \{\%, \%\}, \{\{, \}\},
    },
    literate=
        {\{\%}{{\textcolor{blue}{\{\%}}}{1}
        {\%\}}{{\textcolor{blue}{\%\}}}}{1}
        {\{\{}{{\textcolor{blue}{\{\{~}}}{1}
        {\}\}}{{\textcolor{blue}{~\}\}}}}{1},
    sensitive=true,
    frame=single, 
    framesep=1em, 
    xleftmargin=1em, 
    xrightmargin=1em, 
    morekeywords={video_height,video_width,task_prompt,sub_task_list, current_task, advice_,  drag_start_width, drag_start_height, drag_end_width, drag_end_height, center_width, center_height, operation_type,  operation_value, is_last_action_in_subsession, website, domain, subdomain },
}

Below are the prompt templates sent at different stages, the pipeline controller fills these \textcolor{blue}{\{\{variables\}\}} in them according to the context.

Planning phase prompt template:
\begin{lstlisting}[language=json]
You are familiar with the Linux operating system. 
You can see a computer screen with height: {{ video_height }}, width: {{ video_width }}, and the current task is "{{ task_prompt }}", you need to give a plan to accomplish this goal.
Please output your plan in json format, e.g. my task is to search the web for "What's the deal with the Wheat Field Circle?", the steps to disassemble this task are:
```json 
[
    {"action_type": "PlanAction", "element": "Open web browser."},
    {"action_type": "PlanAction", "element": "Search in your browser for \"What's the deal with the Wheat Field Circle?\""},
    {"action_type": "PlanAction", "element": "Open the first search result"},
    {"action_type": "PlanAction", "element": "Browse the content of the page"},
    {"action_type": "PlanAction", "element": "Answer the question \"What's the deal with the Wheat Field Circle?\" according to the content."}
]
```

Another example, my task is "Write a brief paragraph about artificial intelligence in a notebook", the steps to disassemble this task are:
```json
[
    {"action_type": "PlanAction", "element": "Open Notebook"},
    {"action_type": "PlanAction", "element": "Write a brief paragraph about AI in the notebook"}
]
```
{% if advice_ %}
Here are some suggestions for making a plan: {{ advice_ }}
{% endif %}
Now, your current task is "{{ task_prompt }}", give the disassembly steps of the task based on the state of the existing screen image.
\end{lstlisting}

Acting phase sends prompt:

\begin{lstlisting}[language=json]
You're very familiar with the Linux operating system and UI operations. Now you need to use the Linux operating system to complete a mission. 
Your goal now is to manipulate a computer screen, video width: {{ video_width }}, video height: {{ video_height }}, the overall mission is: "{{ task_prompt }}".

We have developed an implementation plan for this overall mission:
{% for item in sub_task_list %}
    {{ loop.index }}. {{ item }}
{% endfor %}

The current subtask is "{{ current_task }}".
You can use the mouse and keyboard, the optional actions are:
```json
[
    {"action_type":"MouseAction","mouse_action_type":"click","mouse_button":"left","mouse_position":{"width":int,"height":int} },
    {"action_type":"MouseAction","mouse_action_type":"double_click","mouse_button":"left","mouse_position":{"width":int,"height":int} },
    {"action_type":"MouseAction","mouse_action_type":"scroll_up","scroll_repeat":int},
    {"action_type":"MouseAction","mouse_action_type":"scroll_down","scroll_repeat":int},
    {"action_type":"MouseAction","mouse_action_type":"move","mouse_position":{"width":int,"height":int} },
    {"action_type":"MouseAction","mouse_action_type":"drag","mouse_button":"left","mouse_position":{"width":int,"height":int} },
    {"action_type":"KeyboardAction","keyboard_action_type":"press","keyboard_key":"KeyName in keysymdef"},
    {"action_type":"KeyboardAction","keyboard_action_type":"press","keyboard_key":"Ctrl+A"},
    {"action_type":"KeyboardAction","keyboard_action_type":"text","keyboard_text": "Hello, world!"},
    {"action_type":"WaitAction","wait_time":float}
]
```
Where the mouse position is relative to the top-left corner of the screen, and the keyboard keys are described in [keysymdef.h].


Please make output execution actions, please format them in json, e.g. 
My plan is to click the Start button, it's on the left bottom corner, so my action will be:
```json 
[
    {"action_type":"MouseAction","mouse_action_type":"click","mouse_button":"left","mouse_position":{"width":10,"height":760} }
]
```

Another example, my plan is to open Notepad and I see Mousepad app on the screen, so my action will be:
```json
[
    {"action_type":"MouseAction","mouse_action_type":"double_click","mouse_button":"left","mouse_position":{"width":60,"height":135} }
]
```

{% if advice_ %}
Here are some suggestions for performing this subtask: "{{ advice_ }}".
{% endif %}
The current subtask is "{{ current_task }}", please give the detailed next actions based on the state of the existing screen image.
\end{lstlisting}

Reflecting phase sends prompt:
\begin{lstlisting}[language=json]
You're very familiar with the Linux operating system and UI operations.
Your current goal is to act as a reward model to judge whether or not this image meets the goal, video width: {{ video_width }}, video height: {{ video_height }}, the overall mission is: "{{ task_prompt }}".

We have developed an implementation plan for this overall mission:
{% for item in sub_task_list %}
    {{ loop.index }}. {{ item }}
{% endfor %}

Now the current subtask is: "{{ current_task }}".
Please describe whether or not this image meets the current subtask, please answer json format:
Here are a few options, if you think the current subtask is done well, then output this:
```json  {"action_type":"EvaluateSubTaskAction", "situation": "sub_task_success"} ```
The mission will go on.

If you think the current subtask is not done well, need to retry, then output this:
```json  {"action_type":"EvaluateSubTaskAction", "situation": "need_retry", "advice": ""I don't think you're clicking in the right place."} ```
You can give some suggestions for implementation improvements in the "advice" field.

If you feel that the whole plan does not match the current situation and you need to reformulate the implementation plan, please output:
```json  {"action_type":"EvaluateSubTaskAction", "situation": "need_reformulate", "advice": "I think the current plan is not suitable for the current situation, because the system does not have .... installed"} ```
You can give some suggestions for reformulating the plan in the "advice" field.

Please surround the json output with the symbols "```json" and "```".
The current goal is: "{{ task_prompt }}", please describe whether or not this image meets the goal in json format? And whether or not our mission can continue.
\end{lstlisting}

\section{Construction and Processing for the COCO, Widget Captions, and Mind2Web Datasets}\label{appendix:data_construction}

To enhance the model localization capabilities, we use three additional datasets to perform a phased blending with the ScreenAgent Dataset, converting them to keyboard and mouse actions that are consistent with those of the ScreenAgent. The converting operations we performed on each dataset are described below:

\subsection{COCO \& Widget Captions Dataset}

The COCO and Widget Captions are two objective detection datasets. Each sample contains the image, the caption of an object in the image, and the bounding box of this object. We therefore convert this into a mouse click and drag action that requires the model to click on the center of the object or drag to draw a box from the top left corner to the bottom right corner. We use the following answer templates to construct these two tasks:

\begin{lstlisting}[language=json]
My plan is to click {{ task_prompt }}, so my action will be:
```json 
[
    {"action_type":"MouseAction","mouse_action_type":"click","mouse_button":"left","mouse_position":{"width":{{ center_width }},"height":{{ center_height }}}}
]
```
\end{lstlisting}

\begin{lstlisting}[language=json]
My plan is to drag draw a box of {{ task_prompt }}, so my action will be:
```json 
[
    {"action_type":"MouseAction","mouse_action_type":"move","mouse_position":{"width":{{ drag_start_width }},"height":{{ drag_start_height }}}},
    {"action_type":"MouseAction","mouse_action_type":"drag","mouse_button":"left","mouse_position":{"width":{{ drag_end_width }},"height":{{ drag_end_height }}}}
]
```

\end{lstlisting}

\subsection{Mind2Web Dataset}

Mind2Web consists of sessions that consist of multiple steps to accomplish a task on one website. Each session consists of multiple actions, with screenshots of the web page before and after each action.
There are three types of actions in Mind2Web: CLICK, SELECT, and TYPE.
We use the following answer template to transform the Mind2Web dataset according to the same action definition as ScreenAgent.
\begin{lstlisting}[language=json]
To finish "{{ task_prompt }}", I need to finish the current_task"{{ current_task }}" by this action:
{% if operation_type == 'CLICK' or operation_type == 'SELECT' %}
```json 
[
    {"action_type":"MouseAction","mouse_action_type":"click","mouse_button":"left","mouse_position":{"width":{{ center_width }},"height":{{ center_height }}}}
]
```
{% elif operation_type == 'TYPE' %}
```json 
[
    {"action_type":"MouseAction","mouse_action_type":"click","mouse_button":"left","mouse_position":{"width":{{ center_width }},"height":{{ center_height }}}},
    {"action_type":"KeyboardAction","keyboard_action_type":"text","keyboard_text":"{{ operation_value }}"}{% if is_last_action_in_subsession %},
    {"action_type":"KeyboardAction","keyboard_action_type":"press","keyboard_key":"Enter"}{% endif %}
]
```
{% endif %}

\end{lstlisting}

In addition, we have designed a planning template to enhance the ScreenAgent model's planning ability in web scenarios:
\begin{lstlisting}[language=json]
I can see a {{ website }} page about {{ domain }} {{ subdomain }}, and I'm now targeting {{ task_prompt }}.
Based on the screen I'm seeing I've set up some detailed plans for this goal:
```json 
[
{% for item in sub_task_list %}
    {"action_type":"PlanAction", "element":"{{ item }}"}{% if not loop.last %},{% endif %}
{% endfor %}
]
```
\end{lstlisting}

Agents can schedule multiple actions when actions can be done in the same page without jumping to another page.

\twocolumn
\section{Details of ScreenAgent Dataset }\label{appendix:dataset_detail}

We conducted some statistical analysis of the dataset from various perspectives.
As illustrated in \cref{fig:Screen Agent Dataset_(b)}, the dataset encompasses five types of actions, with Mouse action comprising the majority, excluding Plan action and Evaluate action. This aligns with real-world scenarios, where humans predominantly use the mouse for computer interactions. 

\begin{figure}[h]
    \centering
        \subfigure[Task prompt token number distribution]{
            \includegraphics[width=0.2\textwidth]{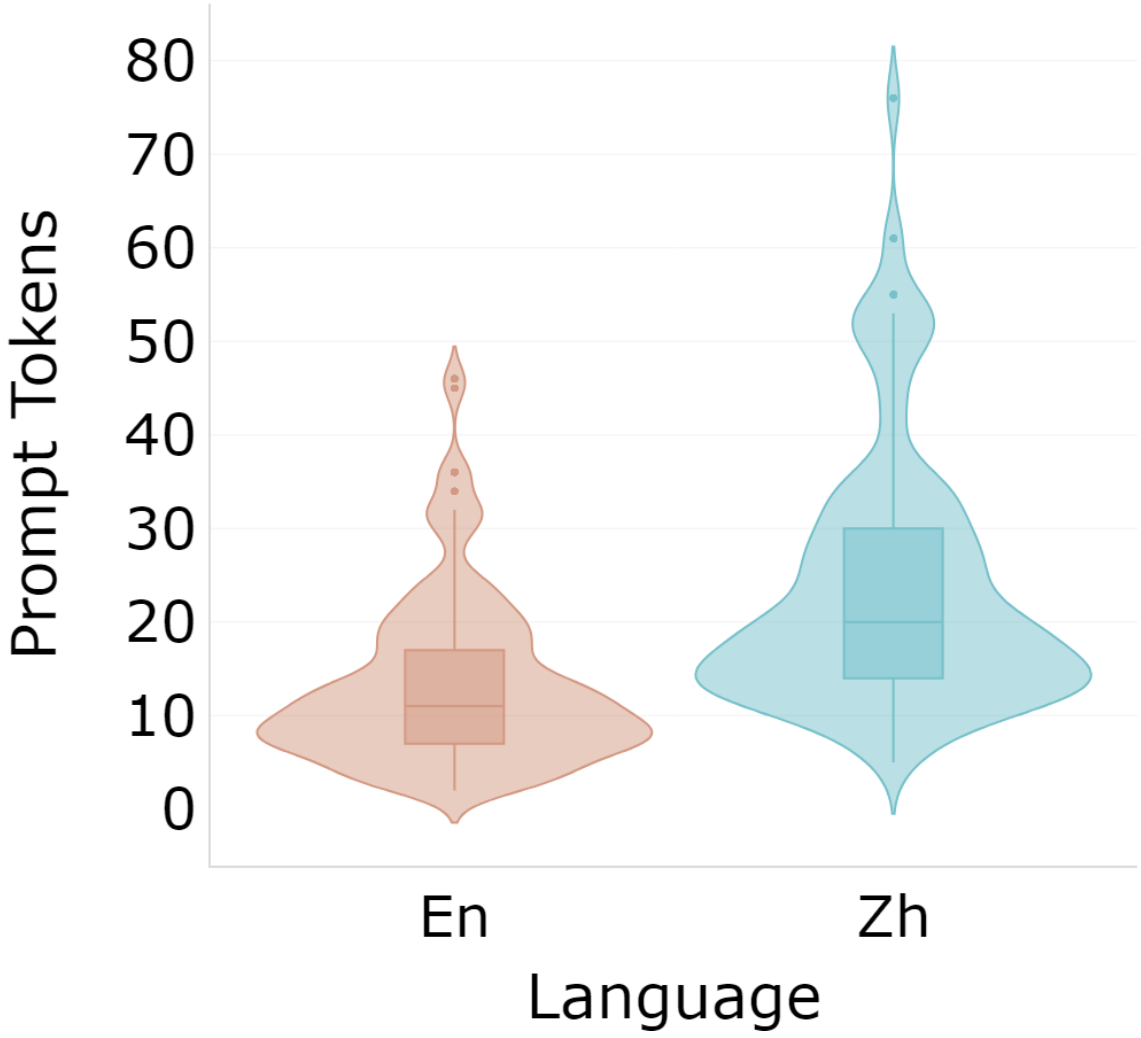}
            \label{fig:App_SAD_(b)}
            }
        \subfigure[Plan element number distribution]{
            \includegraphics[width=0.2\textwidth]{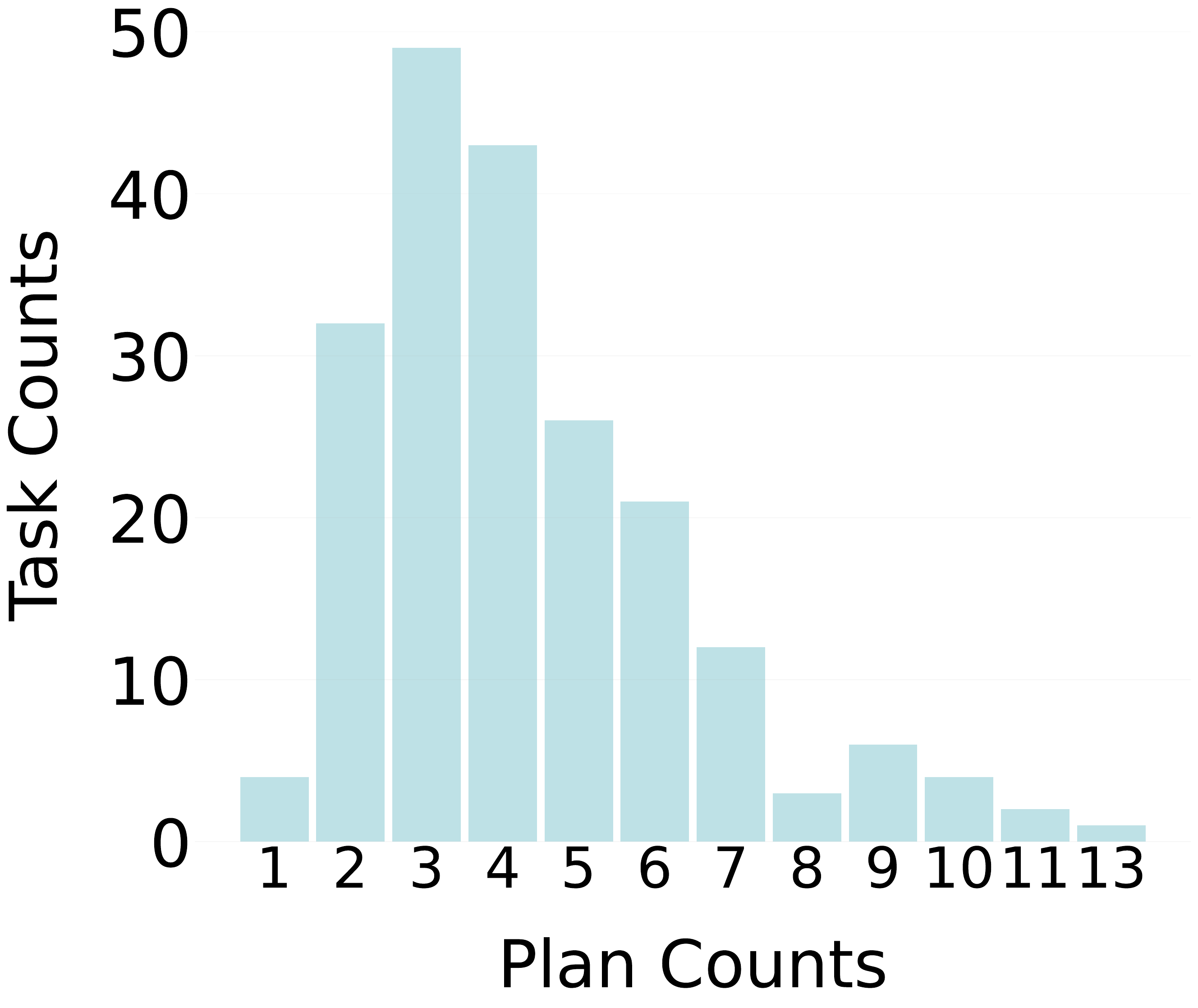}
            \label{fig:App_SAD_(c)}
            }
    \caption{Some statistical information of the ScreenAgent training set.}
    \label{fig:App_ScreenAgentDataset}
\end{figure}

The task prompt represents the overall objective provided by the user. \cref{fig:App_SAD_(b)} and \cref{App_tab:App_token} display the statistical information regarding the number of tokens in the task prompts.
Chinese will consume more tokens than English due to the encoding efficiency of the tokenizer.

The \cref{fig:App_SAD_(c)} illustrates the distribution of the number of subtask plan elements required in the training set to complete the entire task. The most complex tasks demand up to 13 steps in planning. Approximately 60\% of the tasks require a plan consisting of 3 to 5 steps, and the average number of steps in the plans for these tasks is 4.

The statistics regarding the number of actions in subtasks are also included in \cref{App_tab:App_token}. This implies that, on average, 1.5 control actions are required in one interaction with the screen.

\begin{table}[h]
    \centering
    \renewcommand{\arraystretch}{1.2}
    \scalebox{0.78}{
        \begin{tabular}{c|ccc}
          \toprule
          \textbf{Type}    &\textbf{\# Average }& \textbf{\# Max} & \textbf{\# Min} \\ \hline
          \textbf{Task prompt tokens(En)}     & 13.2  & 46  & 2    \\
          \textbf{Task prompt tokens(Zh)}     & 23.8  & 76  & 5   \\
          \textbf{Chosen response tokens(En)} & 97.1  & 779 & 19 \\
          \textbf{Chosen response tokens(Zh)} & 129.9 & 845 & 27 \\
          \textbf{Actions in sub-task}        & 1.5   & 18  & 1 \\
          \bottomrule
        \end{tabular}
    }
    \caption{The statistics about tokens in the ScreenAgent training set.}
    \label{App_tab:App_token}
    \vspace{-1em}
\end{table}

\section{CC-Score}\label{appendix:CC-Score}
Consider two action sequences, namely, the label action sequence $L$ and the pred action sequence $P$. We denote the label action sequence as \( L = \{l_1, l_2, ..., l_n\} \) and the prediction action sequence as \( P = \{p_1, p_2, ..., p_m\} \). Define a score matrix \( S \) as an \( n \times m \) matrix, where \( S_{ij} \) represents the similarity score between action \( l_i \) and action \( p_j \).

A possible alignment is a sequence \( C = \{(c_1, d_1), (c_2, d_2), ..., (c_k, d_k)\} \), where each element \( (c_i, d_i) \) is a pair of actions such that \( c_i \in L \cup \{\text{None}\} \) and \( d_i \in P \cup \{\text{None}\} \). Each pair \( (c_i, d_i) \) either consists of corresponding actions from \( L \) and \( P \), or a combination of an action from either \( L \) or \( P \) with a null value (None). Importantly, this sequence must adhere to the constraint that the order of non-null actions in \( L \) and \( P \) within the alignment is preserved as in their original sequences. This means that if \( (c_i, d_i) \) and \( (c_j, d_j) \) are two pairs in \( C \) where \( c_i, c_j \neq \text{None} \) and \( d_i, d_j \neq \text{None} \), and if \( i < j \), then \( c_i \) must precede \( c_j \) in \( L \) and \( d_i \) must precede \( d_j \) in \( P \). This constraint ensures that the alignment respects the sequential nature and integrity of the original action sequences.

For a given alignment \( C \), its score is the sum of the similarity scores for all matched pairs of actions in the alignment, that is, \( \sum_{(i, j) \in C} S_{ij} \). 
We choose the alignment with the highest score as the best matching alignment \( C^* \).

Finally, the CC-Score for the prediction and label action sequence is calculated as:
$$ \text{CC-Score}(L,P) =  \frac{1}{|L|} \sum_{(i, j) \in C^*} S_{ij} $$

\section{More Generated Samples of ScreenAgent}\label{appendix:sample_display}

To more concretely demonstrate ScreenAgent's performance in accomplishing the task and the ability of the process in the three phases of planning, action, and reflection, we provide additional samples in the following pages.

\begin{figure*}[p]
    \centering
    \includegraphics[width=1\textwidth]{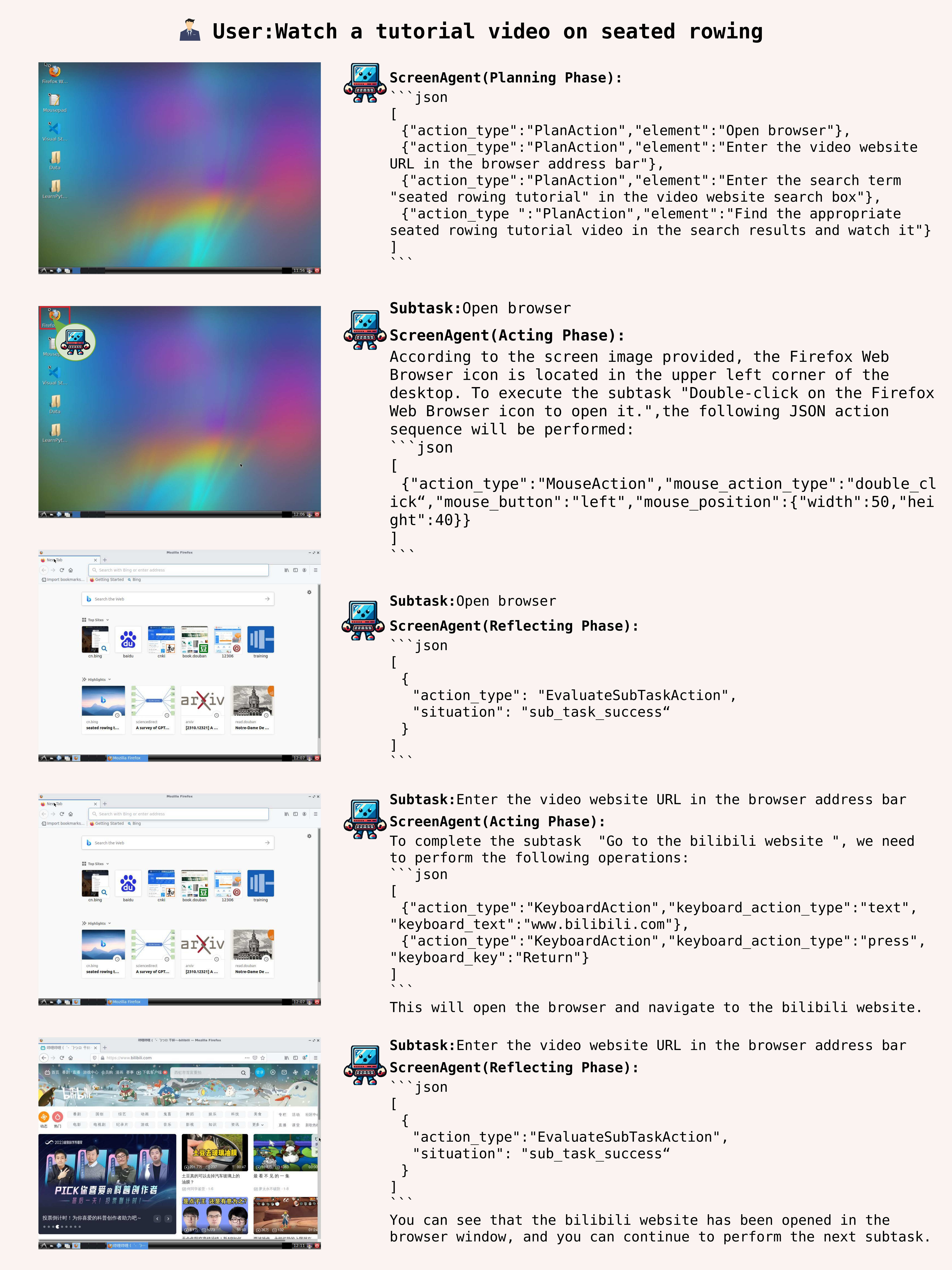}
    \caption{An example of "Watch a tutorial video on seated rowing".}
    \label{fig:appendix_case_1}
\end{figure*}

\begin{figure*}[p]
    \centering
    \includegraphics[width=1\textwidth]{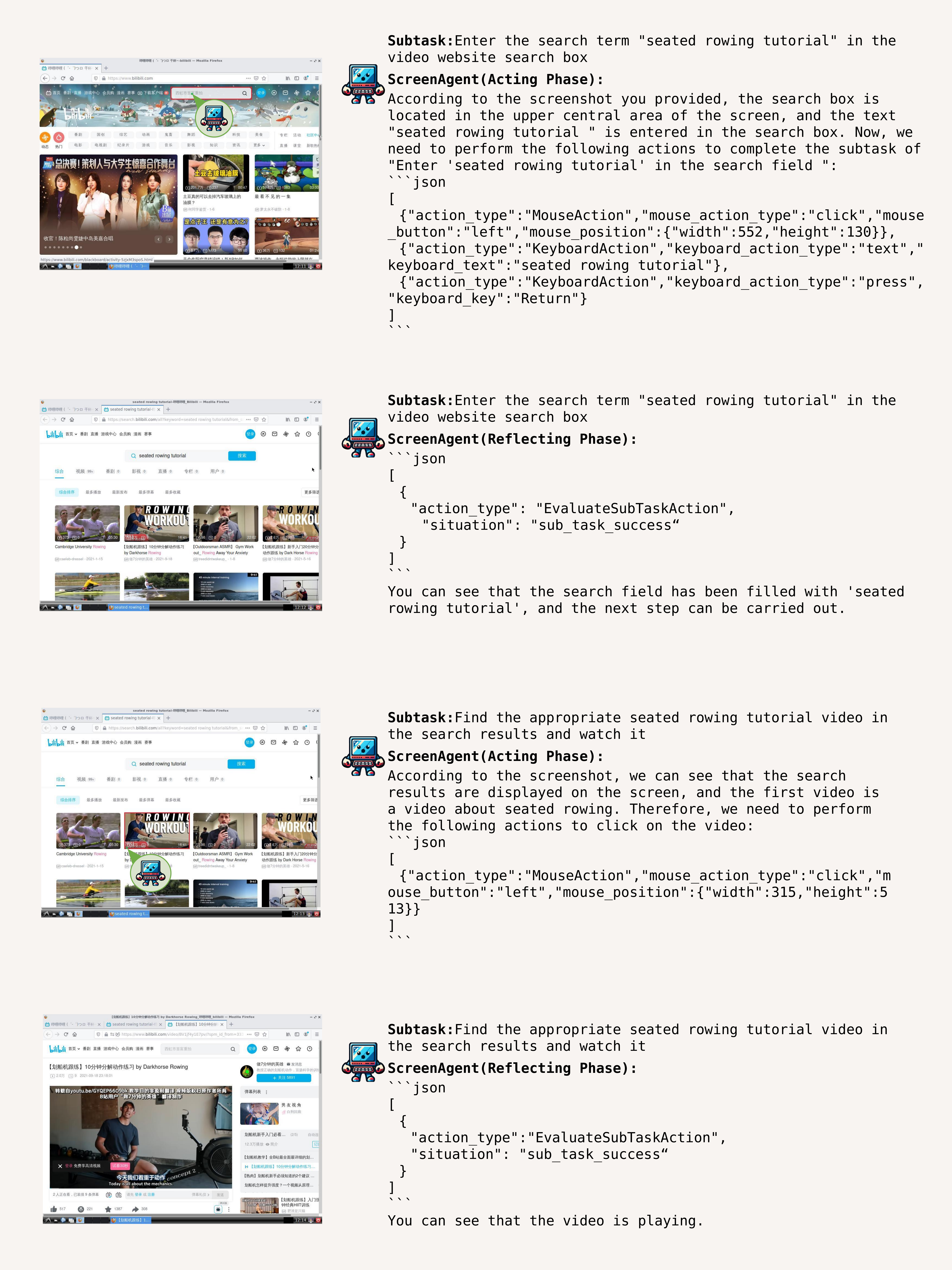}
    \caption{Continue the example of "Watch a tutorial video on seated rowing".}
    \label{fig:appendix_case_1_continue}
\end{figure*}

\begin{figure*}[p]
    \centering
    \includegraphics[width=1\textwidth]{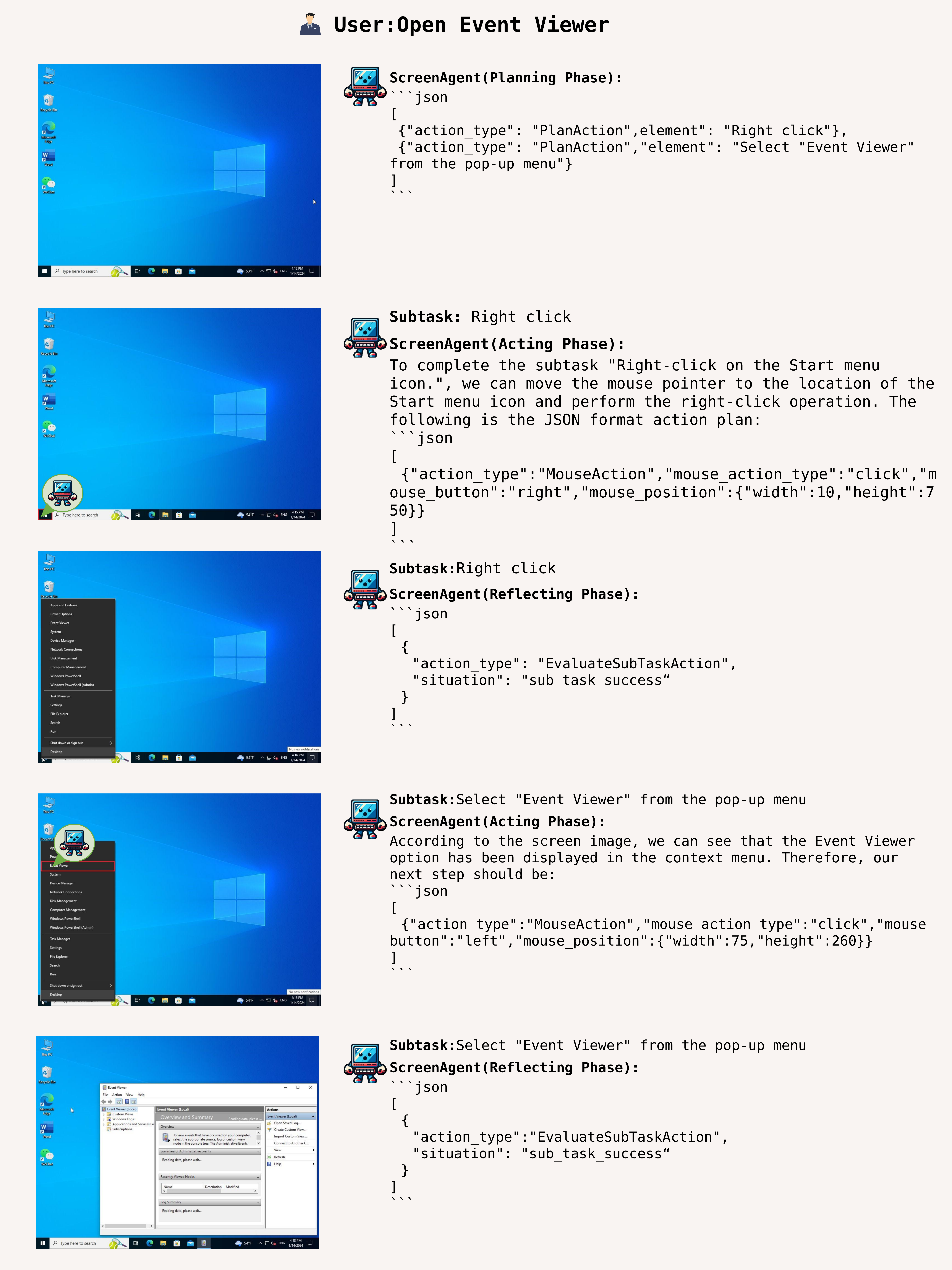}
    \caption{An example of a complete task in the test set.}
\end{figure*}

\begin{figure*}[p]
    \centering
    \includegraphics[width=1\textwidth]{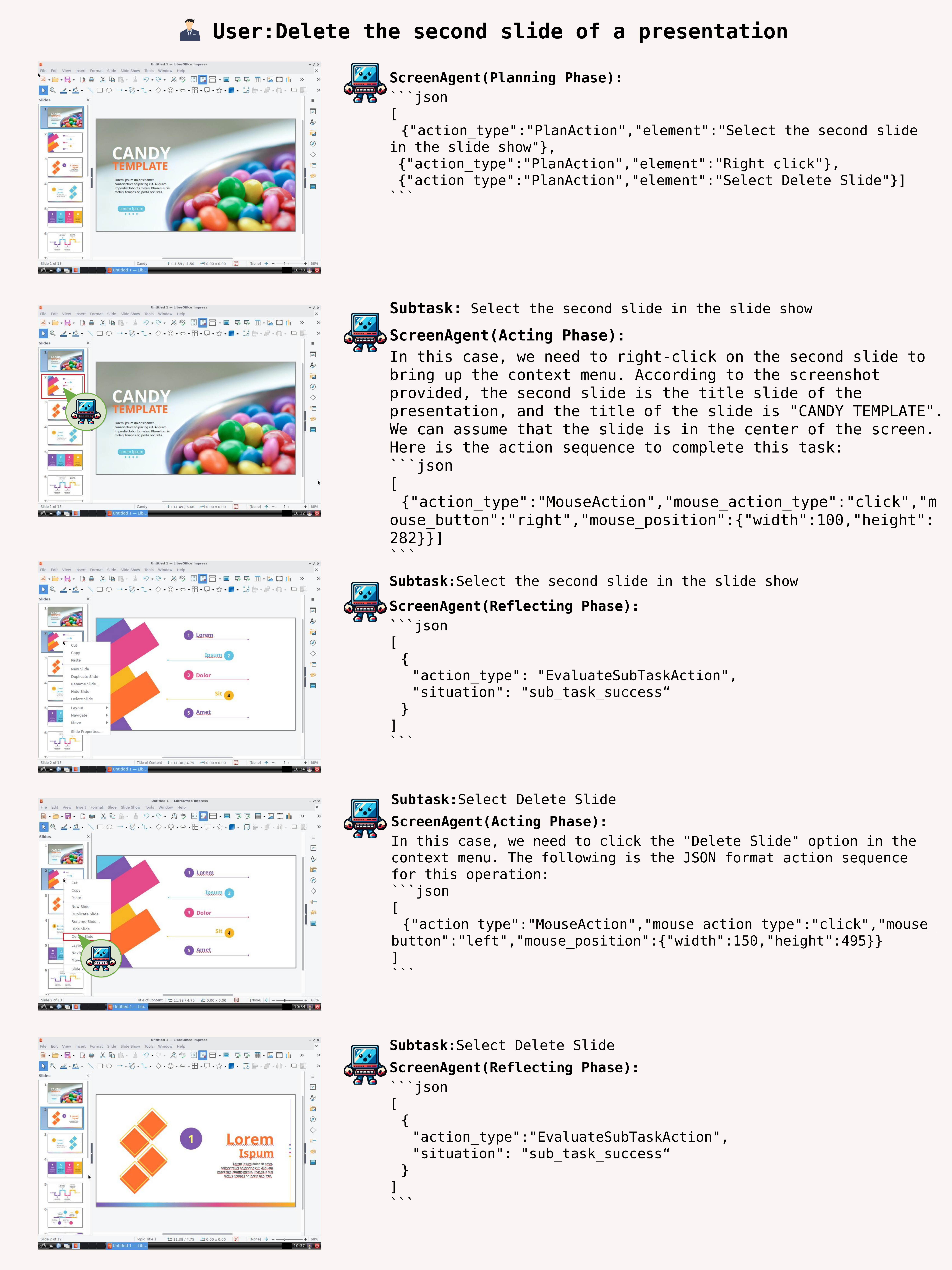}
    \caption{An example of "Delete the second slide of a presentation".}
\end{figure*}

\begin{figure*}[p]
    \centering
    \includegraphics[width=1\textwidth]{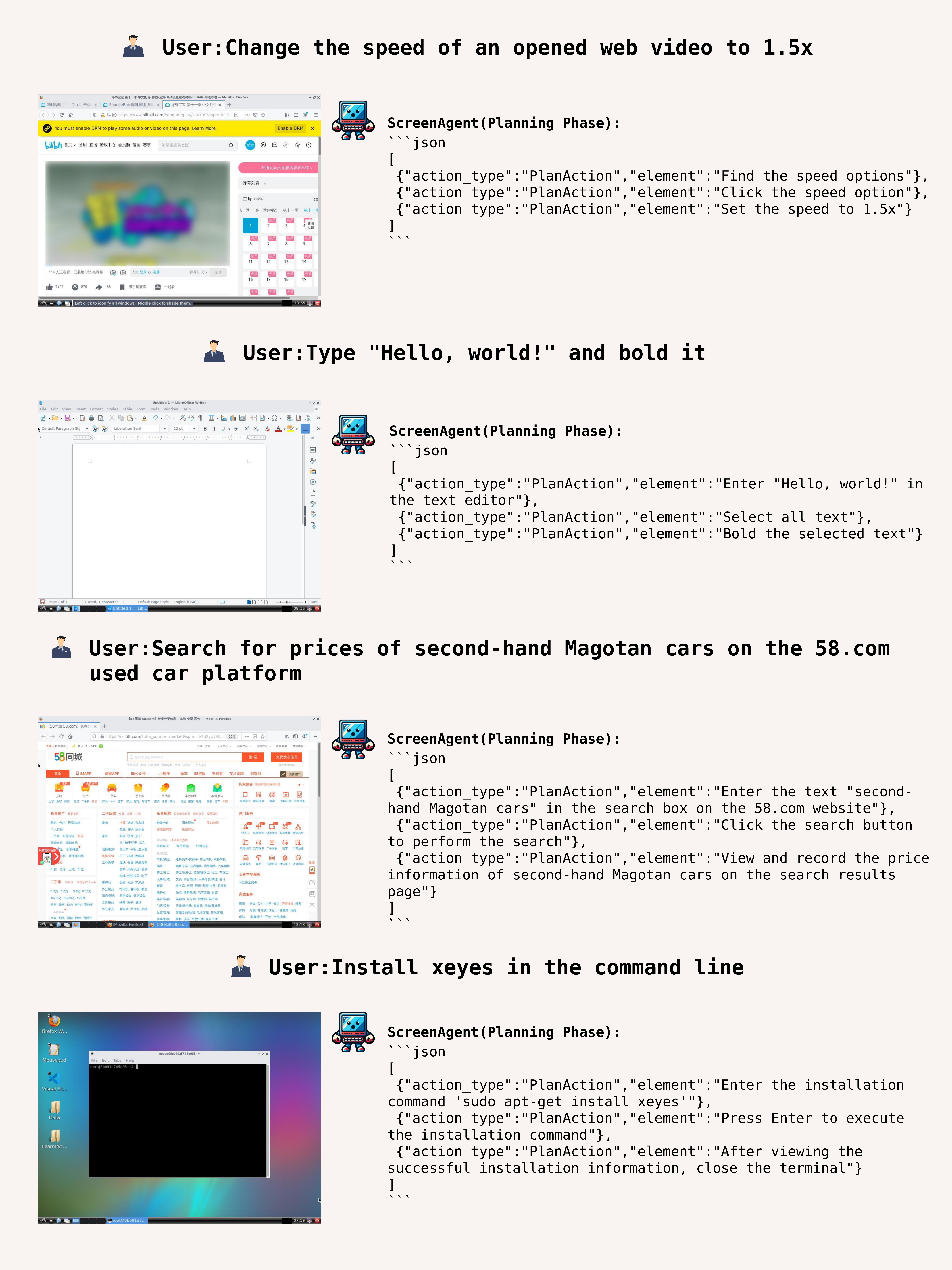}
    \caption{Some examples of the planning stage.}
\end{figure*}

\begin{figure*}[p]
    \centering
    \includegraphics[width=1\textwidth]{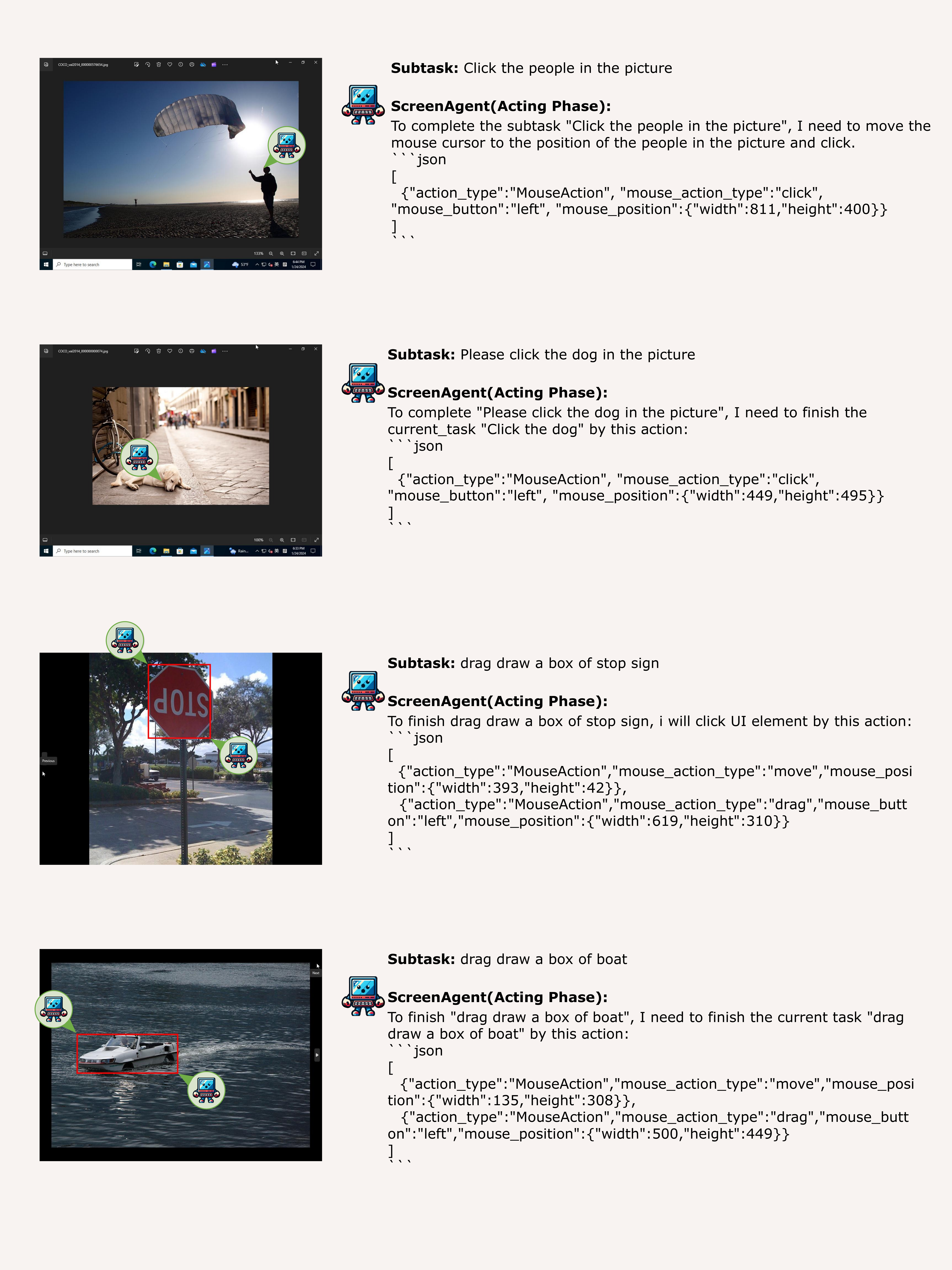}
    \caption{Tested on natural images from the COCO dataset.}
\end{figure*}

\begin{figure*}[p]
    \centering
    \includegraphics[width=1\textwidth]{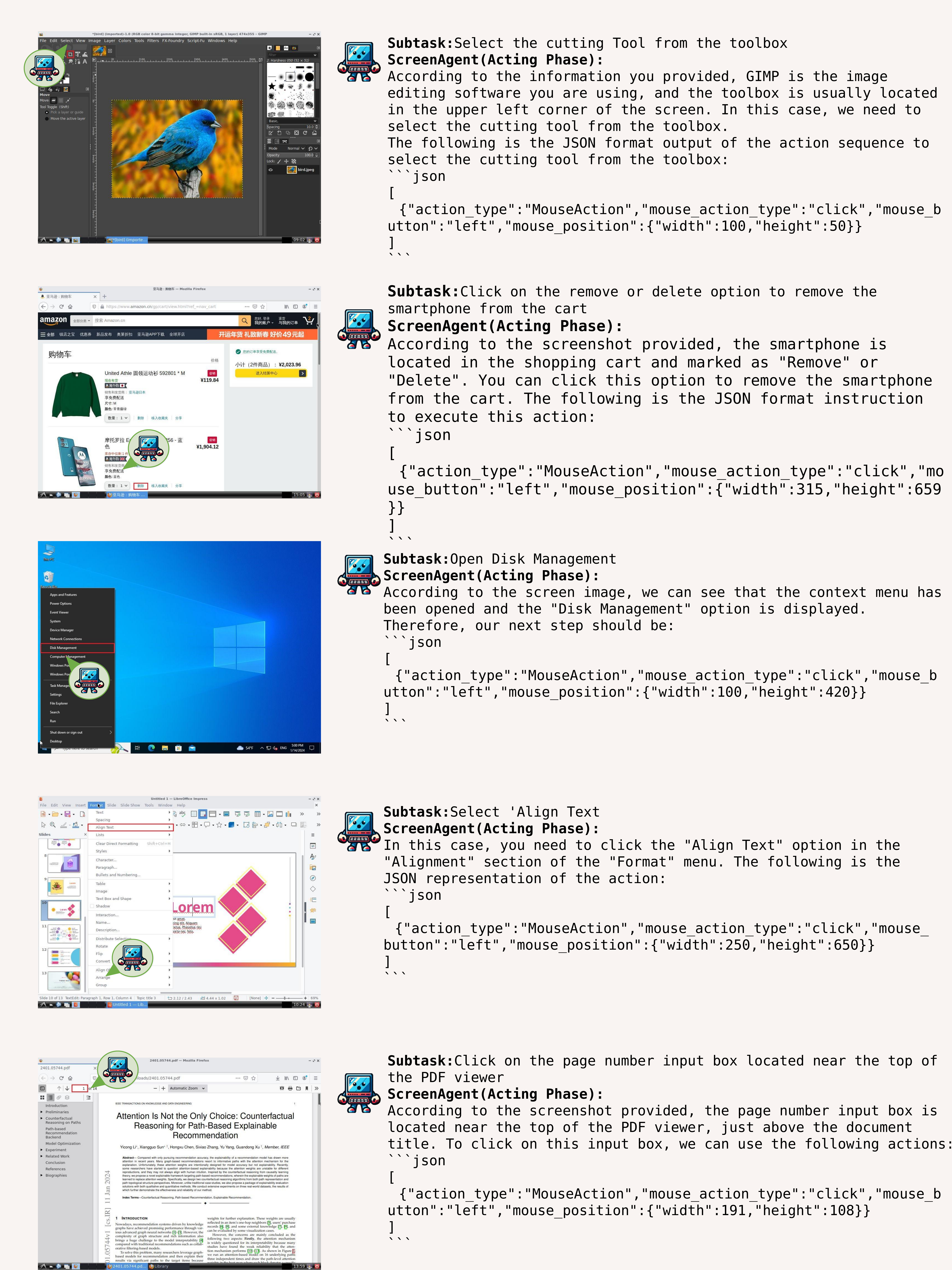}
    \caption{Some cases of failure. The golden label clickable areas are indicated by red bounding boxes.}
\end{figure*}

\section{ScreenAgent Training Configurations}
\cref{tab:hyper-parameters} shows the hyper-parameters we used to train the ScreenAgent model.

\begin{table}[h]
    \centering
    \begin{tabular}{cc}
        \toprule
        Hyper-parameters & Fine-tune \\
        \midrule
        Total steps & 6258 \\
        Warmup step ratio & 0.02 \\
        Learning rate & 1e-5 \\
        Learning rate decay style & Consine \\
        Batch size & 8 \\
        Weight decay & 0.05 \\
        Adam $\epsilon$ & 1e-8 \\
        Adam $\beta$ & (0.9, 0.95) \\
        Gradient clipping & 0.1 \\
        \bottomrule
    \end{tabular}
    \caption{ScreenAgent model training hyper-parameters}
    \label{tab:hyper-parameters}
\end{table}

\end{document}